\documentclass[preprint2]{aastex}

\shorttitle{Short Title Goes Here}
\shortauthors{Somebody et al.}
\begin{document}
%% --------------------------------------------------------
%% --------------------------------------------------------
%% --------------------------------------------------------
\title{16 Years of RXTE Monitoring of Five Anomalous X-ray Pulsars}
\author{Rim~Dib\altaffilmark{1} and
        Victoria~M.~Kaspi\altaffilmark{2}}
\affil{Department of Physics, McGill University,
                 Montreal, QC H3A~2T8}
\altaffiltext{1}{rim@physics.mcgill.ca}
\altaffiltext{2}{vkaspi@physics.mcgill.ca}
%% --------------------------------------------------------
%% --------------------------------------------------------
%% --------------------------------------------------------
%% --------------------------------------------------------
\begin{abstract}
%% --------------------------------------------------------
%% --------------------------------------------------------
%% --------------------------------------------------------
%% --------------------------------------------------------
We present a summary of the long-term evolution of various properties
of the five non-transient Anomalous X-ray Pulsars (AXPs)
1E~1841$-$045, RXS~J170849.0$-$400910, 1E~2259+586, 4U~0142+61, and
1E~1048.1$-$5937, regularly monitored with {\emph{RXTE}} from 1996 to
2012. We focus on three properties of these sources: the evolution of
the timing, pulsed flux, and pulse profile. We report several new
timing anomalies and radiative events, including a putative
anti-glitch seen in 1E~2259+586 in 2009, and a second epoch of very
large spin-down rate fluctuations in 1E~1048.1$-$5937 following a
large flux outburst. We compile the properties of the 11 glitches and
4 glitch candidates observed from these 5 AXPs between 1996 and 2012.
Overall, these monitoring observations reveal several apparent
patterns in the behavior of this sample of AXPs: large radiative
changes in AXPs (including long-lived flux enhancements, short bursts,
and pulse profile changes) are rare, occurring typically only every
few years per source; large radiative changes are almost always
accompanied by some form of timing anomaly, usually a spin-up glitch;
only 20--30\% of timing anomalies are accompanied by any form of
radiative change. We find that AXP radiative behavior at the times of
radiatively loud glitches is sufficiently similar to suggest common
physical origins. The similarity in glitch properties when comparing
radiatively loud and radiatively silent glitches in AXPs suggests a
common physical origin in the stellar interior.  Finally, the overall 
similarity of AXP and radio pulsar glitches  suggests
a common physical origin for both phenomena.
\end{abstract}
%% --------------------------------------------------------
%% --------------------------------------------------------
%% --------------------------------------------------------
%% --------------------------------------------------------
%% --------------------------------------------------------
%% --------------------------------------------------------
%% --------------------------------------------------------
%% --------------------------------------------------------
%% A small citation guide:
%% \citep{jon90} (Jones et al. 1990) [no comma] 
%% \cite(t){jon90} Jones et al. (1990) [no comma]
%%
%% \citep*{jon90} (Jones, Baker, and Williams 1990) [one arg, no comma]
%% \citet*{jon90} Jones, Baker, and Williams (1990) [one arg]
%%
%% \citealp(jon90) Jones et al. 1990  [no comma]
%% \citealt{jon90} Jones et al. 1990  [no comma, been using this]
%%
%% \citealp*{jon90} Jones, Baker, and Williams 1990 [all and no comma,this]
%% \citealt*{jon90} Jones, Baker, and Williams 1990 [all and no comma]
%% --------------------------------------------------------
%% --------------------------------------------------------
%% --------------------------------------------------------
%% --------------------------------------------------------
%% ---------- P A P E R -- S T A R T S -- H E R E ---------
%% --------------------------------------------------------
%% --------------------------------------------------------
%% --------------------------------------------------------
%% --------------------------------------------------------

\section{Introduction}

Prior to the 1995 December launch of NASA's {\it Rossi X-ray Timing
Explorer (RXTE)}, there existed a major puzzle surrounding two
apparently distinct class of high-energy sources, Soft Gamma Repeaters
(SGRs) and Anomalous X-ray Pulsars (AXPs).  At the time, three SGRs
were known and were classified as such due to their distinctive
repeating soft gamma-ray bursts.  The famous `March 5' event of 1979
\citep{mgi+79}, involving SGR 0526$-$66 in the Large Magellanic Cloud
and the release of over $10^{44}$~erg in just a few minutes,
demonstrated the astonishing potential these objects have for
explosive energy releases.  The source of the energy for the bursts
and this giant flare was unknown, but was proposed to be the decay of
an enormous $>10^{14}-10^{15}$~G internal magnetic field in a young
neutron star -- the so-called ``magnetar'' model
\citep{td93a,td95,td96a}.

Meanwhile, roughly half-a-dozen AXPs had been identified by
\cite{vtv95} as `anomalous' X-ray pulsars, spectrally distinct from
conventional accreting pulsars, and lying all in the Galactic Plane,
with one in a supernova remnant.  Their spin-down luminosities failed
by orders of magnitude to account for their apparently stable X-ray
luminosities.  For this reason, they were generally believed to be
some strange form of low-mass X-ray binary, although no evidence for
binary companions was seen \citep[e.g.][]{ms95,bs96,bsss98}.  Indeed
the first {\it RXTE} observations of AXPs were done intending to
search for Doppler shifts due to binary motion \citep{mis98}.
\cite{td96a} however, noted some similarities between the AXPs and the
SGRs, and suggested that both are magnetically powered isolated young
neutron stars.

The post-{\it RXTE} picture of SGRs and AXPs has evolved considerably
and indeed following intensive monitoring campaigns, such as that
described in this paper, these objects appear no longer particularly
`anomalous', though certainly still remarkable. In particular, the
discovery of X-ray pulsations in two SGRs during relatively quiescent
phases, along with the direct measument of spin-down in those objects
\citep{kds+98,ksh+99,hlk+99}, as quantitatively predicted in the
magnetar model, provided compelling evidence that SGRs are indeed
magnetars. Subsequently, the {\it RXTE} discovery of SGR-like bursts
in two AXPs \citep{gkw02,kgw+03}, consistent again with the
expectations of the magnetar model as described by \cite{td96a},
argued strongly for the identification of both classes of objects as
magnetars. The latter discovery was a result of a systematic AXP
monitoring program that also demonstrated the great timing stability
of some AXPs \citep{kcs99}, that AXPs exhibit spin-up glitches
\citep{klc00}, and a variety of other interesting AXP phenomenology
that is generally understandable in the magnetar framework
\citep{gk02,dis+03,wkt+04,gkw04,gk04,gkw06,dkg07,dkg08,dkg09,gdk11}.
While there are still some who argue that the quiescent X-ray emission
of both classes is accretion-driven \citep[e.g.][]{eeea09,aec11},
practically all existing accretion models still demand
magnetar-strength fields to power the observed bursts and giant
flares. Regardless, the concensus today, based largely on {\it RXTE}
observations like those reported on here, is that SGRs and AXPs are
one in the same class of object, with the SGRs the more active of the
family, but with clear transition objects between the two classes
\citep[e.g.][]{kgc+01,kkm+03,ier+10}.  In other words, a continuum of
behaviors between AXPs and SGRs has been observed such that their
class definitions have been heavily blurred and the very names `AXP'
and `SGR' seem synonymous. Nevertheless, in this paper we continue to
refer to the targets as AXPs, mainly for consistency with past
publications based on subsets of the data, but clearly noting that
such nomenclature is somewhat out of date.

Much of today's understanding of AXPs in particular comes from the
long-term systematic {\it RXTE} monitoring program of five AXPs.  This
program used short snapshot observations of the five bright,
persistent AXPs taken regularly every few weeks in order to maintain
full pulse phase coherence.  This allowed the source's spin parameters
to be measured with high precision, enabling sensitivity to glitches,
and providing pulsed flux and pulse profile monitoring, in addition to
sensitivity to bursting behavior. This paper presents a complete
analysis of all {\it RXTE} data for the five AXP monitoring targets,
which are listed along with their basic properties in
Table~\ref{table-sources}.  We present here a systematic and uniform analysis of
data from the commencement of regular monitoring in 1998 (and even, in
some cases, including data prior to that) and up to the final AXP
observations made just before {\it RXTE} was shut off in early 2012.
In total, we have analyzed 3202 individual {\it RXTE} observations of
the targets, a total of 10 Ms taken over a span of 15.7 years. {\it
RXTE} revealed many previous unknown AXP phenomena and answered many
basic questions about AXPs and magnetars, but as this paper shows, it
also raised many new questions that have significant bearing on our
physical understanding of magnetars.

This paper is structured as follows. In \S\ref{sec:overview}, we
present a quick overview of the five AXPs monitored in this project.
In \S\ref{sec:observations} we describe the {\emph{RXTE}}
observations of our targets. In \S\ref{sec:analysis} we
summarize the three kinds of analysis performed: timing, pulsed flux,
and pulse profile. In \S\ref{sec:results}, we detail our results
for each source.  In \S\ref{sec:discussion}, we consider the
behavior of each source -- as well as the collective behaviors of all
the targets -- and discuss our findings and their implications for the
physical nature of these objects.

%% --------------------------------------------------------
%% --------------------------------------------------------
%% --------------------------------------------------------
%% --------------------------------------------------------

\section{Overview and Previous History of the Sources}
\label{sec:overview}

\subsection{1E~1841$-$045}

1E~1841$-$045 is a 7.8-s AXP located in supernova remnant Kes~73 \citep{vg97}. It
was observed by {\emph{RXTE}} roughly twice per month since 1999~February.

A study of a handful of archival X-ray observations of 1E~1841$-$045 spanning 15 years 
suggested the presence of deviations from a linear spin-down which were
initially attributed to timing noise \citep{vg97}.

%% Don't reference the glitch table directly as \ref{TBIG), 
%% otherwise it will appear before Table 1 (TOBS)!
An analysis of the 1997$-$2006 {\emph{RXTE}} monitoring observations
of this source revealed that glitches had occured in 2002, 2003, and
2006 \citep{dkg08}. It also revealed a stable pulsed flux in the
2$-$20~keV band, and a stable pulse profile. \cite{zk10} showed that
the phase-averaged flux was stable during the same period of time. A
fourth glitch was reported by \cite{RIM40YEARS} and \cite{RIMASPEN}.
The parameters for the 2nd glitch were revised by \cite{RIMASPEN} and
are restated in \S\ref{sec:1841} of this work. 
%% and in the glitch Table in Section~\ref{sec:results}.

In 2010, 1E~1841$-$045 exhibited 4 episodes of short high-energy
{\emph{Swift}}-detected bursts. Bursts were detected on May~6 2010,
2011 February~9, 2011 June~23, and 2011 July~2 \citep{lkg+11}.
%% [Reference GCN
%% circulars? 10722, 11673, 11684, 12079, 12079, and 12103 to 12105]
The results of the analysis of the {\emph{RXTE}}
observations collected around the times of these bursts,
as well as the remaining results of {\emph{RXTE}} monitoring of
1E~1841$-$045, are presented in \S\ref{sec:1841}.

\subsection{RXS~J170849.0$-$400910}

RXS~J170849.0$-$400910 is an 11-s AXP discovered in 1996 \citep{snt+97}.

%% ((degrees))
Like 1E~1841$-$045, RXS~J170849.0$-$400910 glitches frequently. The
first glitch detected from this source occured in 1999 \citep{klc00},
the second in 2001 \citep{kg03,dis+03}, and the third in 2005
\citep{dkg08, igz+07}. \cite{dkg08} also reported on several glitch
candidates, and reported no significant changes in the pulsed flux
before 2006. These events were called `glitch candidates' because a
4th or 5th order polynomial fit to the same data near glitch epochs
resulted in similar timing residuals to those obtained from a glitch
fit.

During the same time period as the {\emph{RXTE}} monitoring,
RXS~J170849.0$-$400910 was observed sparsely with different X-ray
imaging satellites {\emph{Chandra}}, {\emph{XMM}} and {\emph{Swift}}.
\cite{roz+05}, \cite{cri+07} and \cite{igz+07} analysed the imaging
observations and reported the source's phase-average flux to be
variable. Because the pulsed flux of the source was reported to be
stable, this suggested an anti-correlation between the phase-averaged
flux and the pulsed fraction. \cite{roz+05}, \cite{cri+07} and
\cite{igz+07} also claimed a correlation between the flux variations
and the glitches.  This flux variation and correlation with glitches
has recently been shown to have been spurious, however (Scholz et al.,
submitted.).

{\emph{RXTE}} monitoring results for
RXS~J170849.0$-$400910 are presented in \S\ref{sec:1708}.

\subsection{1E~2259+586}

1E~2259+586 is a 7-s AXP in the supernova remnant CTB~109 \citep{fg81}
and has been studied extensively.  In particular \cite{bs96} reported
fluctuations in the source's flux and spin down on a timescale of
years. {\emph{RXTE}} monitoring of 1E~2259+586 started in 1996.

%% Iwasawa addition
After several years of stability, 1E~2259+586 entered an outburst
phase in 2002 June in which almost every aspect of the emission
suddenly changed: the pulsed and persistent flux, the pulsed fraction,
the timing properties, the spectral properties, and the pulse profile
\citep{kgw+03,wkt+04}.  There is evidence that a similar event
happened in 1990 \citep{ikh92}. The long-term recovery from the 2002
major outburst was studied by \cite{zkd+08}.

A second glitch occured in 2007 and was reported in \cite{RIM40YEARS},
\cite{RIMASPEN}, \cite{RIMTHESIS}, and \cite{ibi12}. In contrast to
the previous glitch, it was radiatively quiet. \cite{ibi12} also
reported two timing anomalies, the second of which is discussed in
\S\ref{sec:2259} along with a summary of the results for this
AXP. Very recently \cite{aks+13} reported on a sudden spin-down event,
an `anti-glitch,' as observed using {\it Swift} after
our {\it RXTE} monitoring ended.

\subsection{4U~0142+61}

4U~0142+61 is an 8.7-s AXP. It was monitored with {\emph{RXTE}} in
1997 and from 2000 to 2012. \cite{mks05} reported on a possible glitch
having occured in 1999, in a data gap. \cite{dkg07} showed that a
glitch may have occured, but that a slow frequency increase cannot be
ruled out. \cite{dkg07} further showed that from 2000 to 2006, the
source's pulsed flux rose by 29$\pm$8\% in the 2$-$10~keV band. 
%% a number obtained after converting the counts/s/PCU to erg/s/cm$^2$.
There were hints that the rise in the pulsed flux was towards the
low-energy end of the band, which was consistent with the hints of
spectral softening reported by \cite{gdk+10} from the analysis of
archival {\emph{XMM}} data.

%% Figures 11 and 12 of http://arxiv.org/pdf/0905.1256v2.pdf
%% show nicely what the frequency did during the active phase.
In 2006 March, 4U~0142+61 entered an active phase. It exhibited six
X-ray bursts, as seen using {\emph{RXTE}}, the last and largest of which
was detected in 2007 February \citep{gdk11}.  During the active phase,
the pulse morphology changed, then slowly recovered, and the frequency
behaved as though it were recovering from a glitch, although the
glitch parameters were difficult to determine because of the pulse
profile changes. The pulsed flux underwent changes as well, but only
for the duration of the observations containing the bursts. The
results of the {\emph{RXTE}} monitoring of 4U~0142+61 before, during,
and after the active phase are presented in \S\ref{sec:0142}.

\subsection{1E~1048.1$-$5937}

1E~1048.1$-$5937 is a 6.5-s AXP. It has a long history of timing
variability and flux variability at many different wavebands (see for
example, \citealt{mer95}, \citealt{mis98}, and \citealt{pkdn00}).

1E~1048.1$-$5937 has been dubbed the ``anomalous'' Anomalous X-ray Pulsar
because of its unique variability behavior. In 2001 and in 2002, the
AXP exhibited two slow-rise pulsed flux flares, with a risetime on a
timescale of weeks, and a decay on a timescale of months \citep{gk04}.
\cite{gkw02} also reported two bursts from the direction of this
source, which happened near the peak of the first pulsed flux flare.
One of the two bursts was accompanied by a short-term pulsed flux
enhancement. Two other bursts were detected from this source at later
dates \citep{gkw06,dkg09}.

Following the flares, in 2003, the pulsar underwent large (factor
$>$~12) changes in its rotational frequency derivative on a timescale
of weeks to months \citep{gk04}, something never before seen in an AXP. It is
unclear whether this timing-related episode and the preceding
radiative flares are related. This was discussed further by
\cite{dkg09}.

\cite{tmt+05} reported on an {\emph{XMM}} observation of
1E~1048.1$-$5937 in 2004, when the phase-averaged flux was lower than
in 2003 but still not back to its 2000 value, and when the pulsed
fraction was higher than in 2003 but still not back to its 2000
value, indicating that the source was still recovering from the
second flare.  They reported an anti-correlation between the total
flux and the pulsed fraction.

Following the above events, from mid-2004 to 2007~March,
1E~1048.1$-$5937 went through a quiet phase. There was little
variation in the spin-down, in the pulsed flux measured by
{\emph{RXTE}}, and in the total flux measured in a handful of X-ray
imaging observations \citep{tgd+08, dkg09}.

%% Don't reference the glitch table directly as \ref{T2BIG), 
%% otherwise it will appear before Table 1 (TOBS)!
In 2007 March, the source reactivated again \citep{tgd+08, dkg09}. The
pulsed flux rose for the third time during the monitoring program,
this time by a factor $\sim$~3 (2$-$10~keV), and the total flux rose
by a factor of $\sim$~7 (2$-$10~keV). This was simultaneous with the
largest AXP glitch observed by {\emph{RXTE}} (see
\S\ref{sec:results}). A re-analysis of the {\emph{RXTE}} data
performed by \cite{dkg09} showed that the previous two flares observed
from the source were also accompanied by timing events. \cite{tgd+08}
analysed imaging observations from before and after the glitch and
derived an anti-correlation between the pulsed fraction and the total
flux. After the initial onset of the outburst, the timing and
radiative parameters slowly recovered. A few months later, the source
started experiencing rapid changes in the frequency derivative for the
second time. An update on these variations is presented in
\S\ref{sec:1048} along with the remaining results of the
1E~1048.1$-$5937 {\emph{RXTE}} monitoring.

%% --------------------------------------------------------
%% --------------------------------------------------------
%% --------------------------------------------------------
%% --------------------------------------------------------

\section{Observations and Time Series Preparation}
\label{sec:observations}

All observations presented here were obtained using the Proportional
Counter Array (PCA) onboard {\em{RXTE}}. The PCA consisted of an
array of five collimated xenon/methane multi-anode proportional
counter units (PCUs) operating in the 2$-$60 keV range, with a total
effective area of approximately 6500~cm$^2$ and a field of view of
$\sim$1$^{\circ}$ FWHM \citep{jsg+96}.

Throughout the monitoring, we used the {\tt GoodXenonwithPropane} and
the {\tt GoodXenon} data modes to observe our sources. Both data modes
record photon arrival times with 1-$\mu$s resolution and bin photon
energies into one of 256 channels. To maximize the signal-to-noise
ratio, we analyzed only those events from the top Xenon layer of each
PCU. The total number of {\emph{RXTE}} observations analyzed for each
source, as well as the dates of the first and the last analyzed
observations, are shown in Table~\ref{TOBS}.

For each observation, we reduced the data to the solar system
barycenter, and then extracted clean time series binned at a
resolution of 1/32~s. For the timing and pulse profile analyses we
extracted time series that included counts from all operational PCUs
in a source-specific energy band that maximizes the signal-to-noise
(see Table~\ref{TOBS}). For the pulsed flux analysis, we extracted time
series in five bands (2$-$4~keV, 4$-$10~keV, 2$-$10~keV, 4$-$20~keV,
and 2$-$20~keV), not including PCU~0 and PCU~1 because of the loss of
their propane layers. 
%% Not all bands are shown on the summary plots.

%% --------------------------------------------------------
%% --------------------------------------------------------
%% --------------------------------------------------------
%% --------------------------------------------------------

\section{Analysis}
\label{sec:analysis}

%% --------------------------------------------------------
%% --------------------------------------------------------
%% --------------------------------------------------------
%% --------------------------------------------------------

\subsection{Timing analysis}
\label{sec:timinganalysis}

For the timing analysis, 
%% photon arrival times at each epoch were
%% adjusted to the solar system barycenter.  Resulting arrival times were
%% binned with 31.25-ms time resolution. Each barycentric binned time
each barycentric binned time series was epoch-folded using an
ephemeris determined iteratively by maintaining phase coherence as we
describe below. When an ephemeris was not available, we folded the
time series using a frequency obtained from a periodogram.  Resulting
pulse profiles, with 64 phase bins (32 bins in the case of
1E~1841$-$045), were cross-correlated in the Fourier domain with a
high signal-to-noise template created by adding phase-aligned
profiles. The cross-correlation returned an average pulse
time-of-arrival (TOA) for each observation corresponding to a fixed
pulse phase. To estimate the uncertainty on each TOA, we added to each
folded profile many realizations of Poisson noise and
re-cross-correlated each time. The phase uncertainty is determined
from the standard deviation of the resulting simulated TOAs
\citep{kcs99}. The pulse phase $\phi$ at any time $t$ can usually be
expressed as a Taylor expansion,

\begin{equation}
\phi(t) = \phi_{0}(t_{0})+\nu_{0}(t-t_{0})+
\frac{1}{2}\dot{\nu_{0}}(t-t_{0})^{2}
+\frac{1}{6}\ddot{\nu_{0}}(t-t_{0})^{3}+{\ldots},
\label{eq:t1}
\end{equation}

\noindent
where $\nu$~$\equiv$~1/$P$ is the pulse frequency,
$\dot{\nu}$~$\equiv$~$d\nu$/$dt$, etc$.$, and subscript ``0'' denotes a
parameter evaluated at the reference epoch $t=t_0$.

Once the TOAs were obtained, we performed two kinds of phase-coherent
timing analyses on four of the AXPs (1E~1841$-$045,
RXS~J170849.0$-$400910, 1E~2259+586, and 4U~0142+61).

In the first type of timing analysis, we considered all stretches of
time uninterrupted by timing discontinuities (see below). We fitted
the TOAs in each time stretch to the above polynomial using the pulsar
timing software package TEMPO\footnote{See
http://www.atnf.csiro.au/research/pulsar/tempo.}. TEMPO returned the
best-fit polynomials coefficients, the phase-coherent timing solution,
and also returned an absolute pulse number for each TOA. The resulting
timing solutions are plotted as red lines in Panel~a of
Figures~\ref{plot-1841}, \ref{plot-1708}, \ref{plot-2259},
and~\ref{plot-0142}, and the corresponding timing residuals are shown
in Panel~c of the same Figures. Timing parameters from these
fits are presented in Tables \ref{table-1841}, \ref{table-1708},
\ref{table-2259}, and \ref{table-0142}.

%% inter glitch important here
In the second type of timing analysis, we divided each inter-glitch 
(see below) stretch of time into many short overlapping segments. The
number of TOAs contained in each segment depended on the cadence at
which the source was observed. To get from one segment to the next, we
shifted by one TOA. For each small segment, we fitted the TOAs to the
above polynomial, allowing a single frequency derivative $\dot{\nu}$.
The best-fit $\nu$ and $\dot{\nu}$ for all small segments are
presented in Panels~a and~b of Figures~\ref{plot-1841} through
\ref{plot-0142}. The horizontal error bars indicate the length of the
individual segments, and the vertical error bars indicate the
uncertainty in $\nu$ and $\dot{\nu}$.

Sudden changes in $\nu$ and $\dot{\nu}$ are called glitches.  Slow
variations in $\dot{\nu}$ are usually grouped under the name `timing
noise.' AXPs sometimes exhibit events in which it is difficult to
determine whether a change in the timing properties occured abruptly
or progressively over a short period of time. This often happens when
the event occurs near a gap in the data.  When the changes in the
timing parameters at the epoch of such an event do not unambiguously
identify it as a glitch, but are sufficiently abrupt (i.e.,
necessitating an ephemeris consisting of a polynomial of degree 4 or 5
to flatten the residuals near the epoch of the event), the
discontinuities were called `glitch candidates'.  The timing
parameters for the glitches and glitch candidates are presented in
Table~\ref{TBIG}.

Since the spin-down of 1E~1048.1$-$5937 was particularly unstable, and
phase coherence could only be maintained for periods of at most
several months at a time \citep{kgc+01,gk04,tgd+08,dkg09}, the timing
analysis of 1E~1048.1$-$5937 was done differently. We broke the list
of TOAs into many overlapping segments varying in length between 3 and
16 weeks depending on the local noise level of the source. For each
segment we used TEMPO to fit the TOAs using Equation~1 and extracted
absolute pulse numbers. We then checked that the pulse numbers of the
observations present in overlapping segments were the same. This gave
us confidence that the two overlapping ephemerides were consistent
with each other and that phase coherence was not lost.

Combining all overlapping segments between two given dates yielded
long time series of absolute pulse number versus TOA. The
uncertainties on the TOAs were converted into fractional uncertainties
in the pulse numbers. For 1E~1048.1$-$5937, because of irregularities
in the spin-down, timing solutions spanning long periods of time
required the use of very high-order polynomials, which tended to
oscillate at the end points of fitted intervals. To eliminate the
oscillations problem, instead of using these polynomials, we used
splines. A spline is a piecewise polynomial function. It consists of
polynomial pieces of degree $n$ (here $n$~=~5) defined between points
called `knots.' The two polynomial pieces adjacent to any knot share a
common value and common derivative values at the knot, through the
derivative of order $n$$-$2 (see \citeauthor{dierckx}, 1975 for more
details about splines). We fit a spline function through each pulse
number time series, weighed by the inverse of the square of the
fractional errors.  To minimize oscillations in the spline due to
noise, we set the spline smoothing parameter to allow the RMS phase
residual obtained after subtracting the spline from the data points to
be twice the average 1$\sigma$ uncertainty in the pulse phase. The
smoothing parameter controls the tradeoff between closeness and
smoothness of fit by varying the polynomial coefficients and the
spacing between the knots. We found the uncertainties on the spline by
adding Gaussian noise to our data points 500 times, with mean equal to
the 1$\sigma$ uncertainty on each data point, fitting each time with a
spline, averaging all the splines, and finding the standard deviation
at each point.

%% The derivative of the spline function gave us the frequency of the
%% pulsar as a function of time (Panel~a of Fig.~\ref{plot-1048}), and
%% the second derivative of the spline gave us the frequency derivative
%% of the pulsar (Panel~b of Fig.~\ref{plot-1048}).

%% Adding a reference to figure 6 here to make it appear before fig7:
The derivative of the spline function gave us the frequency of the
pulsar as a function of time (Panel~a of Fig.~\ref{plot-1048}), and
the second derivative of the spline gave us the frequency derivative
of the pulsar (Panel~b of Fig.~\ref{plot-1048}). The corresponding
timing residuals are shown in Panel~6 of the same Figure. A comparison
of the timing residuals for the five AXPs is shown in
Figure~\ref{plot-res}; see \S\ref{sec:1708} for details.

%% XMM stuff was moved by Vicky here, and the wording was changed a bit.
For 1E 1841$-$045, we supplemented our {\it RXTE} timing data with
two {\emph{XMM}} observations taken on 2002 October~5 and
2002 October~7, made with the EPIC pn camera in large window
mode.  We made use of these data to help solve a phase ambiguity
described in \S\ref{sec:1841} below.  These observations' OBSIDs are
0013340201 and 0013340101 and had integration times
of 6.6 and 6.0 ks, repsectively.
From these data, photons were extracted in a region
of radius 32.5~arcseconds around the source, and photon arrival times
were adjusted to the solar system barycenter. Then, from each
observation, a time series in the 1.8$-$11~keV range with a time
resolution of 47.7~ms was extracted, and a TOA was obtained from the
time series using the method explained above.

%% --------------------------------------------------------
%% --------------------------------------------------------
%% --------------------------------------------------------
%% --------------------------------------------------------

\subsection{Pulsed Flux Analysis}

To determine the pulsed flux for each observation, we removed any
bursts present in the time series, folded the data and extracted
aligned pulse profiles in several energy bands.  For each folded
profile, we calculated the RMS pulsed flux,

\begin{equation}
F_{RMS} =
{\sqrt{2 {\sum_{k=1}^{n}}
(({a_k}^2+{b_k}^2)-({\sigma_{a_k}}^2+{\sigma_{b_k}}^2))}},
\label{eq:f1}
\end{equation}
\noindent where $a_k$ is the $k^{\textrm{\small{th}}}$ even Fourier
component defined as $a_k$ = $\frac{1}{N} {\sum_{i=1}^{N}} {p_i} \cos
{(2\pi k i/N})$, ${\sigma_{a_k}}^2$ is the variance of $a_k$, $b_k$ is
the odd $k^{\textrm{\small{th}}}$ Fourier component defined as $b_k$ =
$\frac{1}{N} {\sum_{i=1}^{N}} {p_i} \sin {(2\pi k i/N})$,
${\sigma_{b_k}}^2$ is the variance of $b_k$, $i$ refers to the phase
bin, $N$ is the total number of phase bins, $p_i$ is the count rate in
the $i^{\textrm{\small{th}}}$ phase bin of the pulse profile, and $n$
is the maximum number of Fourier harmonics used. For each AXP, we made
pulsed flux series with $n$=2 and $n$=6. For all AXPs, both
series had the same behavior with slightly larger scatter when
the larger number of harmonics was included. Some
of the scatter may be due to low-level fluctuations in the pulse
profile; see Panel e of Figures~\ref{plot-1841} through~\ref{plot-1048}. 
The pulsed flux time series for each AXP is
presented in Panel d of Figures~\ref{plot-1841} through
~\ref{plot-1048}.

%% --------------------------------------------------------
%% --------------------------------------------------------
%% --------------------------------------------------------
%% --------------------------------------------------------

\subsection{Pulse Profile Analysis}

For each observation, we folded the data in the same energy band used
for timing using the best-fit frequency found in the timing analysis.
We then cross-correlated the resulting profile with a standard
template in order to obtain phase-aligned profiles. We used 32 phase
bins for the aligned profiles. We then subtracted the respective
averages from each of the aligned profiles and from the template. For
each observation, we then found the scaling factor that minimized the
reduced ${\chi}^2$ of the difference between the scaled profile and
the template. The resulting reduced ${\chi}^2$ values are plotted in
Panel e of Figures~\ref{plot-1841} through \ref{plot-1048}.  These
values are generally close to 1 except near the epochs of some major
outbursts. For each AXP, the pulse profile for a typical observation,
as well as the long-term average pulse profile, are shown in
Figure~\ref{plot-profiles}.

%% --------------------------------------------------------
%% --------------------------------------------------------
%% --------------------------------------------------------
%% --------------------------------------------------------

\subsection{Searching for Bursts}

In addition to the timing, pulsed flux, and pulse profile
analyses, we performed our burst-search routine
%% \footnote{The burst
%% search routine returned several candidate bursts with a significance
%% several orders of magnitude smaller than those reported for the
%% published bursts. We do not report on the analysis of these putative
%% bursts here.} 
introduced in \citet{gkw02} and discussed further in \citet{gkw04} on
all the analysed observations. In short, for each data set, to
determine whether a burst occured in the $i^{\rm th}$ time bin (of
duration 31.25 ms in our analysis), the number of counts in that bin
is compared to a local mean. The local mean is calculated over a
stretch of four pulse periods of data, centered around the time bin
being evaluated. A window of one pulse cycle is also administered so
that counts directly from, and immediately around, the point under
investigation would not contribute to the local mean. Bursts found by
our searching procedure are presented in Table~\ref{table-bursts}.

%% --------------------------------------------------------
%% --------------------------------------------------------
%% --------------------------------------------------------
%% --------------------------------------------------------

\section{Results}
\label{sec:results}

%% --------------------------------------------------------
%% --------------------------------------------------------
%% --------------------------------------------------------
%% --------------------------------------------------------

\subsection{1E~1841$-$045}
\label{sec:1841}

\begin{figure}
\centerline{\includegraphics[scale=0.70]{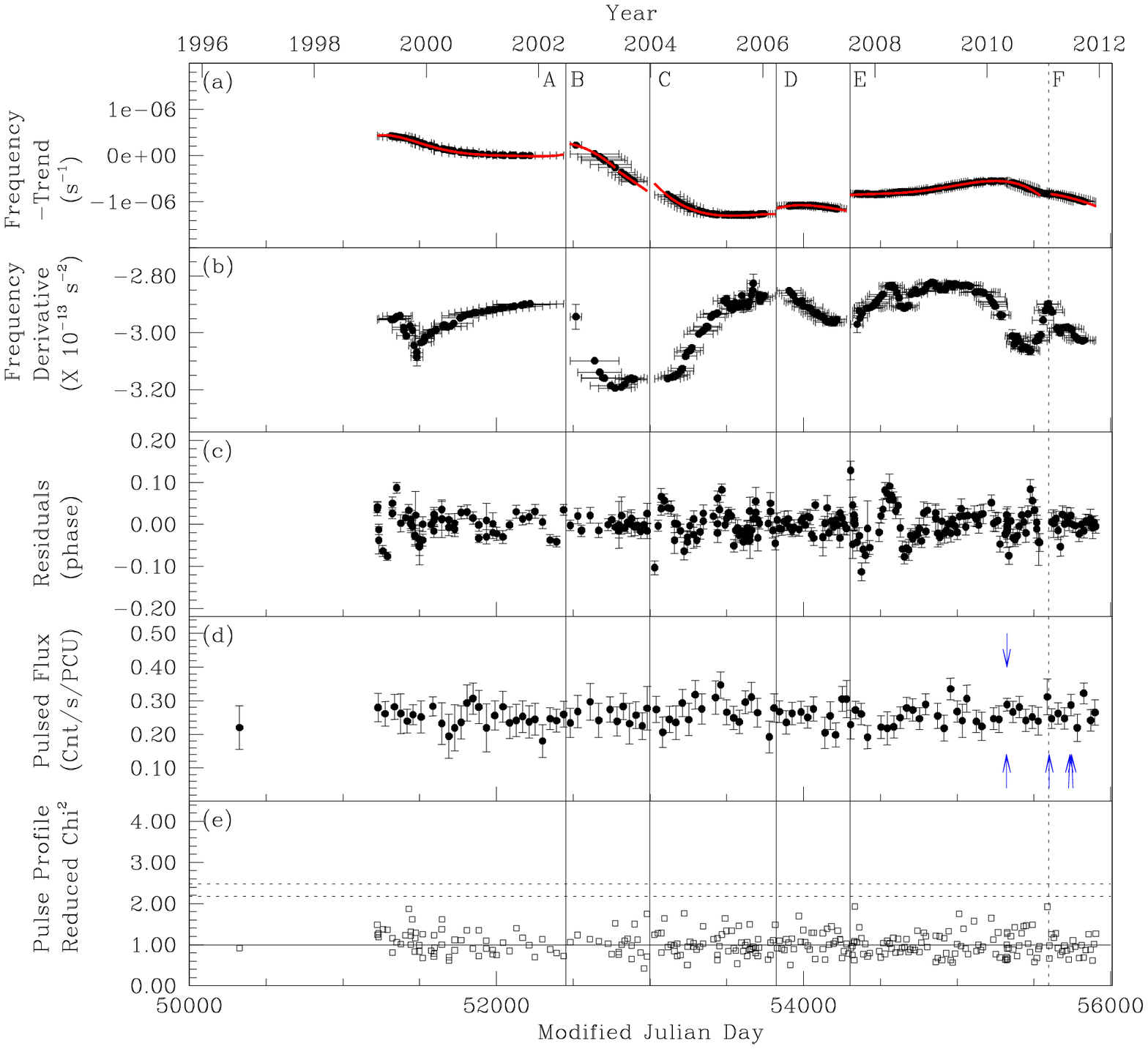}}
\caption
{ %% open caption
\footnotesize
Long-term evolution of the properties of 1E~1841$-$045.  
(a)
Frequency as a function of time with a linear trend in frequency
subtracted.
%% \footnote{(trend:
%% 0.0849040502519462-2.898023384606E-13$\times$86400$\times$(x-52460.0)).}
Red lines: long-term polynomial phase-coherent timing solutions.
Polynomial start and stop times are at glitch epochs, candidate-glitch
epochs, and epochs of other timing discontinuities.
Data points: each data point corresponds to a frequency measurement 
obtained from a phase-coherent timing fit to a small number of TOAs 
(see \S\ref{sec:timinganalysis} for details).
(b) Frequency derivative as a function of time.
Data points: each data point corresponds to a $\dot{\nu}$ measurement
obtained from a phase-coherent timing fit to a small number of TOAs
surrounding the epoch of the data point
(see \S\ref{sec:timinganalysis} for details).
(c) Timing residuals
corresponding to the red lines in the first Panel.
(d) Pulsed flux in the 2$-$20~keV band, grouped in 30-day bins.
The pulsed flux behaved similarly in all five analysed bands.
The four blue arrows pointing upward indicate the location of the
{\emph{Swift}}-detected bursts \citep{lkg+11}. The blue arrow pointing down indicates
the location of the {\emph{RXTE}}-detected burst (see
Table~\ref{table-bursts}).
(e) Reduced ${\chi}^{2}$ statistics for the pulse profiles versus time, calculated after
subtracting the scaled and aligned profiles of the individual 
observations from a high signal-to-noise template for 1E~1841$-$045.
The solid horizontal line indicates a reduced ${\chi}^{2}$ of 1.
The lower dotted line corresponds to the 2$\sigma$ significance level, 
after having taken the number of trials into account.  
The upper dotted line corresponds to the
3$\sigma$ significance level.
%% The solid vertical lines indicate the location of glitches, 
%% and the dotted vertical line indicates the approximate
%% location of the candidate glitch.
Vertical lines indicate timing discontinuities.
Discontinuities marked by solid vertical lines are
glitches and the dotted vertical line indicates
the notable timing discontinuity (see Table~\ref{TBIG}).
\normalsize
\label{plot-1841}
} %% close caption
\end{figure}

A summary of the behavior of 1E~1841$-$045 between 1996 and 2012 as
seen by {\emph{RXTE}} is shown in Figure~\ref{plot-1841}. The
long-term timing parameters of the source are presented in
Table~\ref{table-1841}.

Panels a and b of Figure~\ref{plot-1841} reveal that 1E~1841$-$045 is
a noisy source: the rotational frequency derivative varies
significantly on a timescale of years.  Panel c shows the residuals
following the subtraction of our best-fit timing models, reported in
Table~\ref{table-1841}.

1841$-$045 has exhibited four large glitches since the start of the
monitoring program, marked by the vertical lines in
Figure~\ref{plot-1841} \citep{dkg08,RIM40YEARS}. \cite{dkg08} reported
two possible sets of parameters for the first glitch from
1E~1841$-$045, with the most likely set involving an exponential
recovery. However the addition of two TOAs extracted from {\emph{XMM}}
observations revealed the less likely timing solution to be the
correct one \citep{RIMASPEN, RIMTHESIS}.

%% XMM stuff was removed from here.

There were no significant changes in the pulsed flux in any of the
five studied bands near glitch epochs, nor were there any in
glitch-free intervals. The 2$-$20~keV pulsed flux time series is shown
in Panel~d of Figure~\ref{plot-1841}.  \cite{zk10} showed that the
source's phase-averaged flux is also stable, indicating that the
glitches of 1E~1841$-$045 appear radiatively quiet.

In 2010 and 2011, {\emph{Swift}} detected several episodes of bursts
from 1E~1841$-$045, indicated by the four blue arrows pointing upward
in the lower portion of Panel~d of Figure~\ref{plot-1841}.
\cite{lkg+11} studied the bursting activity with {\emph{Swift}} and
{\emph{Fermi}} and found that it did not have a significant effect on
the persistent flux level of the source. {\emph{RXTE}} observations
show that the pulsed flux of 1E~1841$-$045 appeared featureless around
that time.

However, there were two hints in {\emph{RXTE}} data indicating that
the source was undergoing activity of some kind. First, there was a
small burst (unresolved in a 31-ms time bin) observed from the
direction of 1E~1841$-$045 on 2010 May 7, the day following the first
episode of {\emph{Swift}}-detected bursts (see
Table~\ref{table-bursts}). Note however that in the observation
containing the burst, only a single PCU was on, making the pulse
signal-to-noise ratio too low to allow verification of the presence of
any fluctuations in the pulsed flux within that observation. Also note
that in the {\emph{RXTE}} observation collected on the previous day,
there were no bursts.

The second indication that 1E 1841$-$045 was undergoing activity of
some kind was a timing discontinuity (dotted line in
Figure~\ref{plot-1841}). The following is the sequence of events
surrounding that discontinuity.
%% ((degree))
From 2010 December~8 to 2011 January~16, there were no observations of
the source made with {\emph{RXTE}} because of the angular proximity of
the source to the Sun. The first observation following the data gap
was nominal, with no detected bursts, pulsed flux enhancement, or
significant pulse profile changes. The second {\emph{Swift}}-detected
burst episode then occured on the February~8 and~9. This was followed
again by two seemingly normal {\emph{RXTE}} observations on February~9
and~10. However, the behavior of the timing residual around this time
was peculiar: there was a change in the curvature of the timing
residuals when a long-term polynomial timing solution fit included the
stretch of time around the second burst episode, indicating a change
in one or more timing parameters near that epoch.
%% (())
The change in the curvature of the residuals was not sufficiently abrupt to
warrant calling the event a glitch candidate. Also, because of the
presence of the gap in the data, it is difficult to determine whether
the change in the timing parameters occured during the gap, between
January~16 and February~10, or between February~10 and 23.

%% ((degree)) 
The peak in Panel~b of Figure~\ref{plot-1841} that is indicated by the
dotted vertical line provides another way of seeing the same
phenomenon: each of the data points that are part of the rise and fall
of the peak in Panel~b include TOAs from before and after the data gap
at the same time. Had there been no change in any of the timing
parameters in or near the gap, this peak would not exist. 
%% degree THREE
However, because the data gap makes it impossible to constrain how
fast and when the change in the timing parameters occured, and because
the nearby TOAs can be fitted to an ephemeris consisting of a small
number of frequency derivatives (here only three derivatives), we are
not classifying the event as a glitch candidate.
%% classifying this discontinuity as a glitch candidate.
This event is reported in Table~\ref{TBIG} as a notable timing
discontinuity, along with the glitch parameters obtained from the most
likely glitch epoch.

%%There were no significant {\emph{RXTE}}-detected pulse profile changes
%%for this source (see Fig. \ref{plot-profiles}).
%% %% The pulse profile for a typical observation, as well
%% %% as the long-term average pulse profile for this AXP and for the four
%% %% others are shown for reference in Figure~\ref{plot-profiles}.

%% wording remains
Panel~e of Figure~\ref{plot-1841} shows no significant
{\emph{RXTE}}-detected pulse profile changes for this source. Note
however that this Figure shows the reduced ${\chi}^{2}$ statistics for
individual pulse profiles, and although no significant changes are
seen in this Figure, we believe there are constantly slow low-level
changes, only detectable by summing the pulse profiles over an
extended period of time; see for example Figure~11 of \cite{dkg08}.

%% NOTE ABOUT THE FIGURE ORDER: 1841plot, 1708plot, REZPLOT, 2259plot,
%% 0142plot, 1048plot.

%% --------------------------------------------------------
%% --------------------------------------------------------
%% --------------------------------------------------------
%% --------------------------------------------------------

\subsection{RXS J170849.0$-$400910}
\label{sec:1708}

%% `most dominant fluxtuations''.

\begin{figure}
\centerline{\includegraphics[scale=0.70]{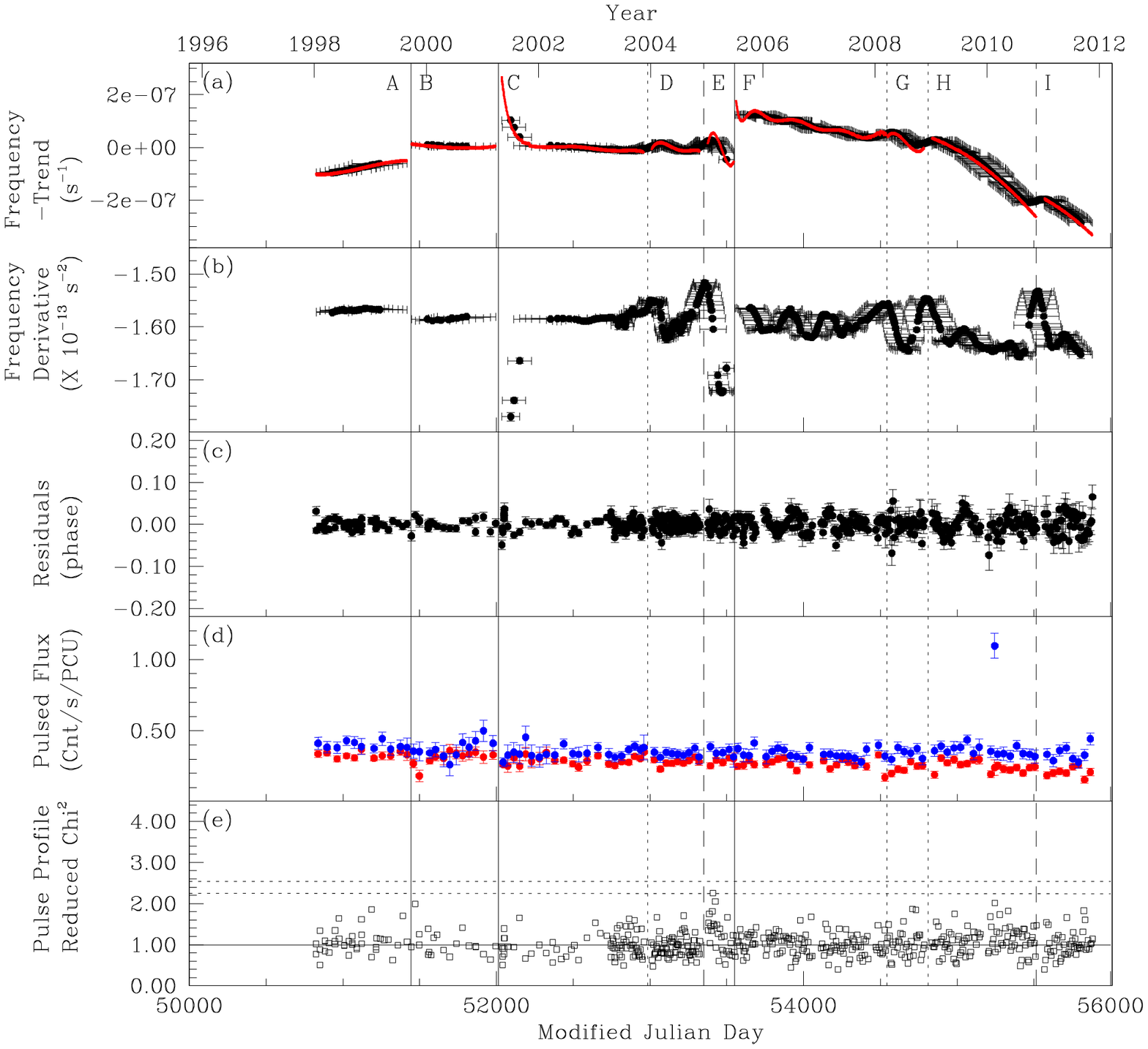}}
\caption
{ %% open caption
\footnotesize
Long-term evolution of the properties of RXS~J170849.0$-$400910.
(a)
Frequency as a function of time with a linear trend in frequency
subtracted.
%%\footnote{(trend:
%%.}
Red lines: long-term polynomial phase-coherent timing solutions.
Polynomial start and stop times are at glitch epochs, candidate-glitch
epochs, and epochs of other timing discontinuities.
Data points: each data point corresponds to a frequency measurement
obtained from a phase-coherent timing fit to a small number of TOAs.
(b) Frequency derivative as a function of time.
Data points: each data point corresponds to a $\dot{\nu}$ measurement
obtained from a phase-coherent timing fit to a small number of TOAs
(see \S\ref{sec:timinganalysis} for details).
(c) Timing residuals
corresponding to the red lines in the first Panel.
(d) Pulsed flux in the 2$-$4~keV band (red points), and in the
4$-$20~keV band (blue points)
grouped in 36-day bins.
(e) Reduced ${\chi}^{2}$ statistics for the pulse profiles versus
time, calculated after
subtracting the scaled and aligned profiles of the individual
observations from a high signal-to-noise template for RXS~J170849.0$-$400910.
The solid horizontal line indicates a reduced ${\chi}^{2}$ of 1.
The lower dotted line corresponds to the 2$\sigma$ significance level,
after having taken the number of trials into account.  
The upper dotted line corresponds to the
3$\sigma$ significance level.
The solid vertical lines indicate the location of glitches,
the dashed vertical lines indicate the locations of glitch candidates, and
the dotted vertical lines indicate the location of other timing discontinuities
(see Table~\ref{TBIG}).
\normalsize
\label{plot-1708}
} %% close caption
\end{figure}

A summary plot of the behavior of RXS J170849.0$-$400910 between 1996
and 2012 as seen by {\emph{RXTE}} is shown in Figure~\ref{plot-1708}.
The long-term timing parameters are presented in
Table~\ref{table-1708}.

%% anomalous #1 mentioned here.
Panels~a and~b of Figure~\ref{plot-1708}, show that RXS
J170849.0$-$400910 underwent timing discontinuities: the solid lines
mark the location of 3 glitches, one of which had an exponential
recovery. Panel~d shows that these glitches, like those of
1E~1841$-$045, showed no detectable pulsed flux variations; the
exception to this being a single anomalous pulsed flux point in the
4$-$20~keV band in early 2010, far from the epochs of any timing
discontinuities. We have studied this anomalous data point in detail
but see no reason to distrust it. In addition to the glitches, there
were many smaller-magnitude timing discontinuities, marked by the
remaining vertical lines, and detected as peaks in Panel~b.

%% israel et al paragraph
Note that \cite{igz+07} analysed a subset of these same data. In
particular, they reported a glitch corresponding to our first glitch
candidate, marked by the first dashed vertical line in
Figure~\ref{plot-1708}. For that event, the reported fit parameters
are similar though not identical to ours (see Table~\ref{TBIG}). They
also reported a glitch coincident with the third solid vertical line
of Figure~\ref{plot-1708}. For that glitch, the reported frequency
jump at the glitch epoch was similar to ours but the jump in frequency
derivative was significantly different. We find that this difference
is due to their inclusion of more post-glitch TOAs when fitting the
glitch.

%% ((degree))

It is possible to compare the degree of abruptess of the changes in
the timing parameters of J170849.0$-$400910 at the epochs of the
various discontinuities by examining the peaks in Panel~b of
Figure~\ref{plot-res}. In this Figure, timing residuals of all AXPs
are shown for selected stretches of time, after the removal of a
long-term trend in frequency. The changes in the curvature of the
residuals are, as expected, most abrupt at the location of the
glitches. This Figure also provides a good way of visually comparing
the amount of timing noise in the various sources by observing the
number of ``wiggles'' in the timing residuals for a given number of
frequency derivatives in the ephemeris used.

%% wording slightly changed by Vicky
Panel~e of Figure~\ref{plot-1708} shows no significant
{\emph{RXTE}}-detected pulse profile changes for this source.
As for 1E~1841$-$045, however, we believe
there could be constant slow low-level changes, only detectable by
summing the pulse profiles over an extended period of time; see for
example Figure~8 of \cite{dkg08}.

%% --------------------------------------------------------
%% --------------------------------------------------------
%% --------------------------------------------------------
%% --------------------------------------------------------

\subsection{1E~2259+586}
\label{sec:2259}

%% Brouillon
%% **
%% seconds AXP has the smallest magnitude nudot, and of the 5 is the
%% one with the smallest magnetic field. Hitory of the source: glitch was
%% reported in 2009 with an exponential recovery. accompanied by an
%% outburst sudden rise in fluxes and bursts, etc. profile changes too.
%% **
%% The results of the timing
%% analysis are presented in Figure XX. Explain the figure. mrntion the
%% overlap number. mention the trne.d remember to say that we used 9 xmm
%% data points in there as well. Remember to say there are two event data
%% data points too. pulsed flux rms method with variance subtraction see
%% for example (other papers0). xmm points blended well.
%% **
%% RIGHT HERE TALK ABOUT ANY DEVIATIONS FROM A LINEAR SPIN-DOWN. looks
%% stable except maybe during the reovery from the 1st glitch. Any overall
%% sstematic
%% deviations?
%% **
%% The first glitch was reported in XX. We report the second glitch here
%% (shown in the long-term figure). A plot of the pre-fit and post-fit
%% timing residuals is shown in Fig. XX. The glitch parameters are shown
%% in the first column of Table XX. Compare the size of this glitch to
%% the size of the previous one.

\begin{figure}
\centerline{\includegraphics[scale=0.70]{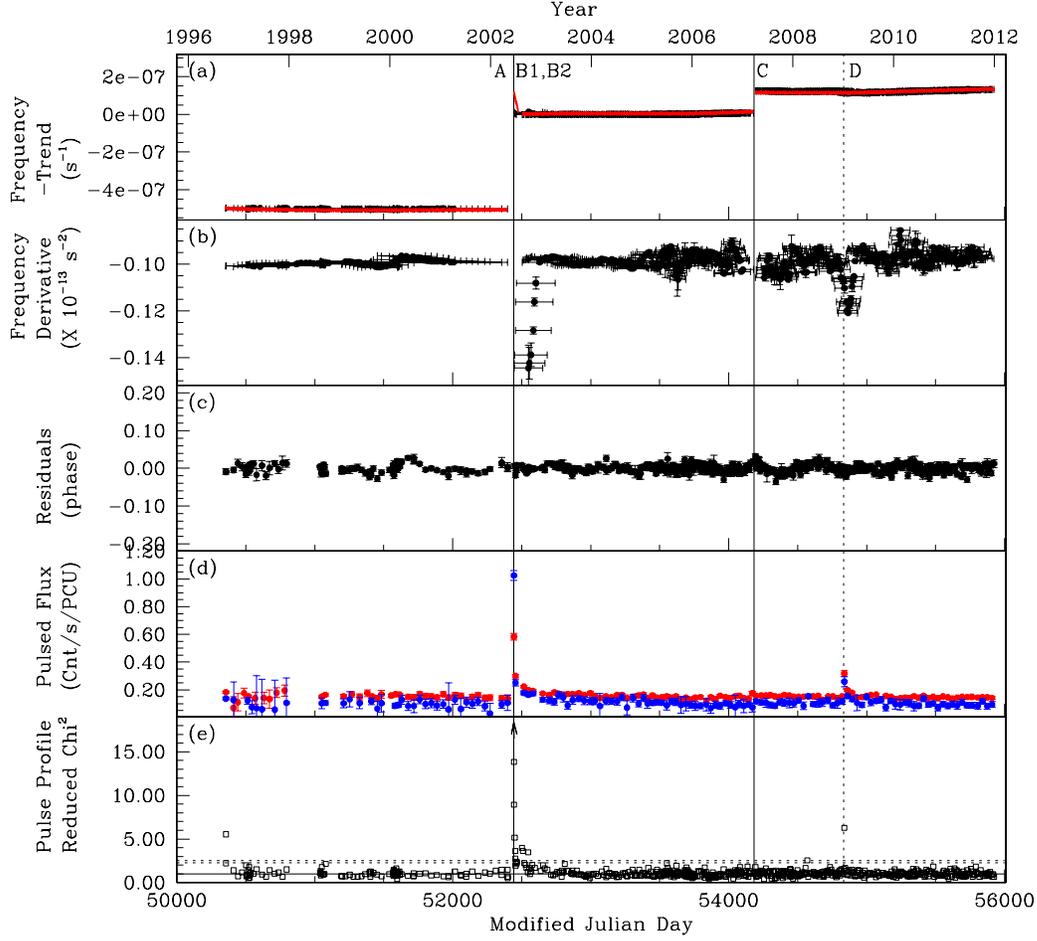}}
\caption
{ %% open caption
\footnotesize
Long-term evolution of the properties of 1E~2259+586.
(a)
Frequency as a function of time with a linear trend in frequency
subtracted.
%%\footnote{(trend:
%%.}
Red lines: long-term polynomial phase-coherent timing solutions.
Red lines start
and stop at glitch epochs, candidate-glitch
epochs, and epochs of timing discontinuities.
The B ephemeris is split into two segments. The first segment ends
at the end of the recovery from the large 2002 glitch.
Data points: each data point corresponds to a frequency measurement
obtained from a phase-coherent timing fit to a small number of TOAs.
(b) Frequency derivative as a function of time.
Data points: each data point corresponds to a $\dot{\nu}$ measurement
obtained from a phase-coherent timing fit to a small number of TOAs
(see \S\ref{sec:timinganalysis} for details)
(c) Timing residuals
corresponding to the red lines in the first Panel.
(d) Pulsed flux in the 2$-$4~keV band (red points), and in the
4$-$20~keV band (blue points)
grouped in 30-day bins.
The first post-glitch observation in 2002 contained bursts \citep{gkw04}.
(e) Reduced ${\chi}^{2}$ statistics for the pulse profiles versus
time, calculated after
subtracting the scaled and aligned profiles of the individual
observations from a high signal-to-noise template for 1E~2259+586.
The solid horizontal line indicates a reduced ${\chi}^{2}$ of 1.
The lower dotted line corresponds to the 2$\sigma$ significance level,
after having taken the number of trials into account.  
The upper dotted line corresponds to the
3$\sigma$ significance level.
The first post-glitch observation in 2002 is off the scale.
The solid vertical lines indicate the location of glitches.
The dotted vertical line indicates the location of a timing
discontinuity; see \S\ref{sec:2259} and Table~\ref{TBIG} for details.
\normalsize
\label{plot-2259}
} %% close caption
\end{figure}

A summary of the behavior of 1E~2259+586 between 1997 and 2012 is
presented in Figure~\ref{plot-2259}. The long-term timing parameters
of the source are presented in Table~\ref{table-2259}.

Panels~a and~b of Figure~\ref{plot-2259}, and more clearly Panel~c of
Figure~\ref{plot-res}, show that 1E~2259+586 exhibits very little
timing noise compared to the other AXPs. As shown by the solid
vertical lines in Figure~\ref{plot-2259}, 1E~2259+586 exhibited two
glitches during our monitoring program.

The recovery of the first of the two glitches could be fit by a
combination of exponentials \citep{kgw+03,wkt+04}.  If the first
post-glitch observation, which contained a large number of bursts, is
excluded, the post-glitch data can also be fit with a simple ephemeris
that includes one frequency derivative, followed by a long-term
ephemeris that contains three frequency derivatives (red lines B1 and
B2 in Panel~a of Figure~\ref{plot-2259}). This glitch was accompanied
by a large enhancement in the pulsed flux (Panel~d of
Figure~\ref{plot-2259}). The highest data point in this Panel is the
average pulsed flux for the observation containing bursts.  Note,
however, that the pulsed flux during that observation fell
monotonically with time; see \cite{kgw+03} and \cite{wkt+04} for
details. The glitch was also accompanied by pulse profile changes
(Panel~e of Figure~\ref{plot-2259}).

%% possible refs \cite{dkg08} \cite{RIM40YEARS} \cite{RIMASPEN}  
%% \cite{RIMTHESIS} \cite{ibi12}
In contrast to the first glitch, the second glitch from this source
was not followed by a recovery, and was radiatively quiet
\citep{dkg08,RIM40YEARS,RIMASPEN,ibi12}.

%% ((degree))
Small but significant changes in the pulsed flux were again seen in
two observations in 2009 January (Panel~d of Figure~\ref{plot-2259})
in all bands within the 2$-$20~keV range. The pulsar exhibited a pulse
profile change during the first of these two observations (Panel~e of
Figure~\ref{plot-2259}). One or more of the timing parameters changed
within 50 days of the anomalous observation.  (This can be most easily
seen in Panel~b of Figure~\ref{plot-2259} and in Panel~c of
Figure~\ref{plot-res}). The TOAs near this event can be fitted both to
a sudden jump in frequency and to a polynomial with several frequency
derivatives. Since the polynomial has only degree $n=3$, the event is
not classified as a glitch candidate, but is reported as a notable
timing discontinuity in Table~\ref{TBIG} because of the associated
radiative changes. Moreover, it is also not classified as a glitch
candidate because it is difficult to determine whether the change in
the timing parameters occured slowly or abruptly. If abruptly, the
epoch of the event can only be narrowed down to within a time period
of 50 days.

This timing discontinuity is notable as when a glitch fit is
attempted, independent of where the glitch epoch is chosen, the jump
in frequency $\Delta\nu$ has a {\emph{negative}} value between
$-$1.00$\times$10$^{-8}$ and $-$1.42$\times$10$^{-8}$ Hz. This fact
was remarked on in \cite{ibi12} as well, although they did not report
the contemporaneous radiative changes, and their reported $\Delta\nu$
is 3.3-$\sigma$ away from ours.

Finally, note that the anomalous ${\chi}^{2}$ near MJD~50357 (Panel~e
of Figure~\ref{plot-2259}) corresponds to the pulse profile of a very
long observation, supporting the idea that there are constant low-level
pulse profile changes that go undetected because of the low
signal-to-noise ratio in most observations.

%% --------------------------------------------------------
%% --------------------------------------------------------
%% --------------------------------------------------------
%% --------------------------------------------------------

\subsection{4U~0142+61}
\label{sec:0142}

\begin{figure}
\centerline{\includegraphics[scale=0.70]{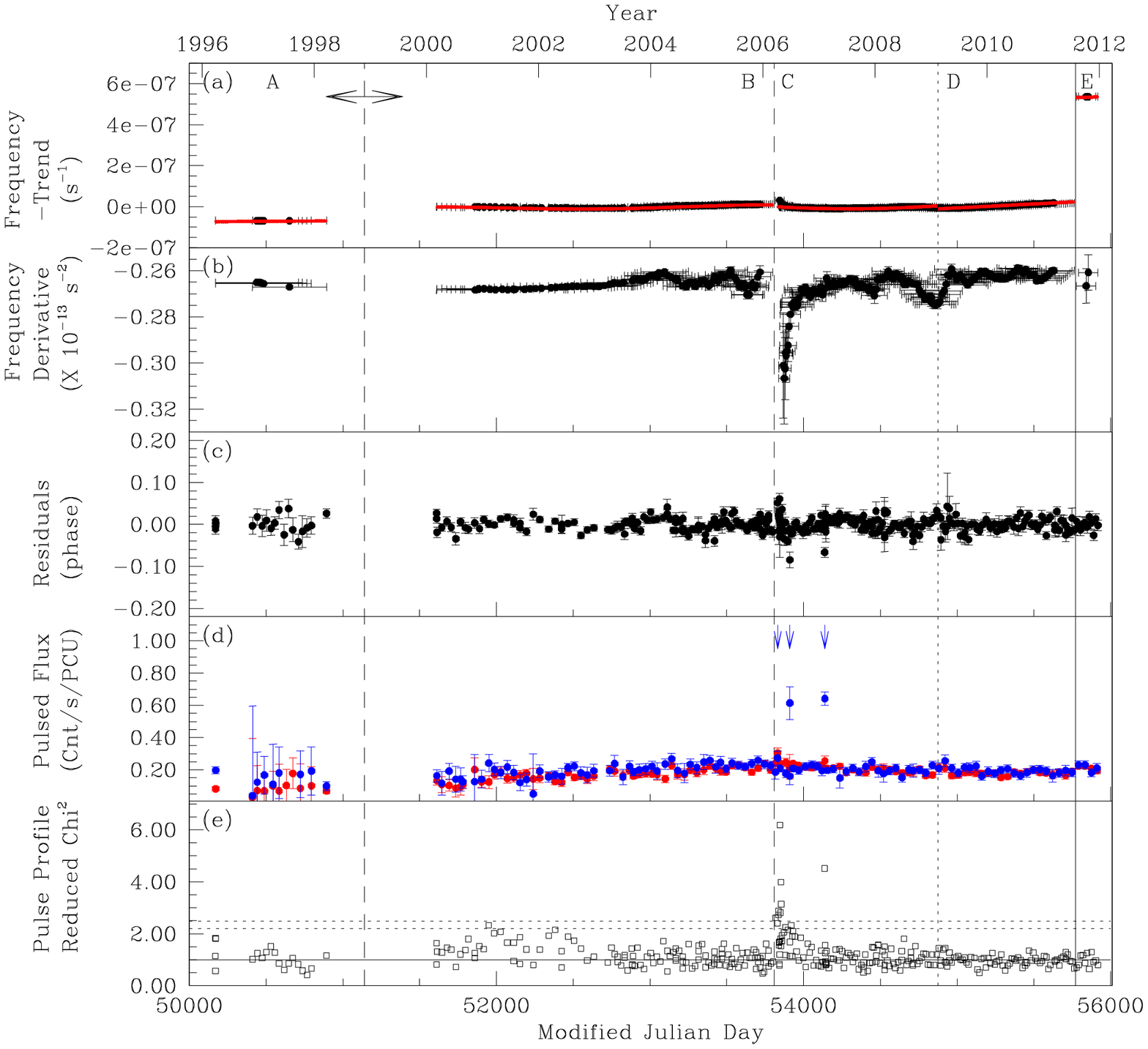}}
\caption
{ %% open caption
\footnotesize
Long-term evolution of the properties of 4U~0142+61.
(a)
Frequency as a function of time with a linear trend in frequency
subtracted.
%% \footnote{(trend:
%% .}
%% WISH I COULD KEEP THE FOLLOWING SENTENCE
%% There are small ripples in the frequency curve that are not visible
%% because of the scale of the plot.
Red lines: long-term polynomial phase-coherent timing solutions.
Polynomial start and stop times are at 
glitch epochs, candidate-glitch
epochs, and epochs of timing discontinuities.
Data points: each data point corresponds to a frequency measurement
obtained from a phase-coherent timing fit to a small number of TOAs.
The double-ended arrow indicates the time period during which the
glitch candidate occured, as reported in \cite{mks05}.
(b) Frequency derivative as a function of time.
Data points: each data point corresponds to a $\dot{\nu}$ measurement
obtained from a phase-coherent timing fit to a small number of TOAs.
(c) Timing residuals
corresponding to the red lines in the first Panel.
(d) Pulsed flux in the 2$-$4~keV band (red points), and in the
4$-$20~keV band (blue points)
grouped in 30-day bins.
The blue arrows indicate
the locations of the {\emph{RXTE}} observations containing bursts.
(e) Reduced ${\chi}^{2}$ statistics for the pulse profiles versus
time, calculated after
subtracting the scaled and aligned profiles of the individual
observations from a high signal-to-noise template for 4U~0142+61.
The solid horizontal line indicates a reduced ${\chi}^{2}$ of 1.
The lower dotted line corresponds to the 2~$\sigma$ significance level,
after having taken the number of trials into account.  
The upper dotted line corresponds to the
3~$\sigma$ significance level.
The solid vertical lines indicate locations of glitches. 
The dashed vertical lines indicate the
locations of glitch candidates.
The dotted vertical lines indicate locations of other timing
discontinuities (see Table~\ref{TBIG}).
\normalsize
\label{plot-0142}
} %% close caption
\end{figure}

A summary of the behavior of 4U~0142+61 between 1996 and 2012 as seen
by {\emph{RXTE}} is shown in Figure~\ref{plot-0142}. The long-term
timing parameters are presented in Table~\ref{table-0142}.

As can be seen from Figure~\ref{plot-0142}, four noteworthy timing events
occured during the X-ray monitoring of 4U~0142+61.

First, there is an offset in the frequency between the red lines
representing ephemeris~A and ephemeris~B in Panel~a of
Figure~\ref{plot-0142}. Based on the comparison of {\emph{RXTE}} and
{\emph{ASCA}} data, \cite{mks05} pointed out that a glitch may have
occured in the gap between the two ephemerides. The event is marked as
a glitch candidate in Figure~\ref{plot-0142} and in Table~\ref{TBIG}.

%% arrow points
The second interesting series of changes occured when the source
entered an active phase in 2006 (second glitch candidate in
Figure~\ref{plot-0142}, timing parameters in Table~\ref{TBIG}). The
pulsar's timing parameters changed, the pulse profile varied, and the
pulsed flux increased locally within the three observations in which
bursts were detected (arrows in Panel~d). The details of this 2006
active phase of 4U~0142+61 were discussed by \cite{gdk11}. 
The timing event associated with this active phase is classified as a
glitch candidate because the claim of a large sudden frequency jump is
based on a single TOA, the first of the active phase, and that TOA may
have been affected by pulse profile changes \citep{gdk11}. If this TOA
is omitted, the initial sudden spin-up is less significant, although
it is clear that a change in $\dot{\nu}$ occured.  Also, even if this
TOA is omitted, extending the post-recovery ephemeris backward in time
makes it look as though an `anti-glitch' occurred \citep{gdk11}.

Next occured a small and possibly slow timing discontinuity in 2009,
most easily seen from Panel~b of Figure~\ref{plot-0142} and from
Panel~d of Figure~\ref{plot-res}. This discontinuity is similar to the
one exhibited by 1E~2259+586 in early 2009 (see
\S\ref{sec:2259}) in that it can be fit both by a local
polynomial of degree $n=3$ in frequency and by a {\emph{negative}}
frequency jump of 1.3(2)$\times$10$^{-8}$~Hz, very similar in size to
that of the 1E~2259+586 event. However, unlike the 2009 discontinuity
in 1E~2259+586, this event did not have contemporaneous radiative
changes.

Finally, a large glitch occured in 2011~July. It was a radiatively
silent glitch: there was at most a statistically marginal increase in
the pulsed flux in all bands (see Fig.~\ref{plot-0142}). Note that we
chose the scale of Panel~a of Figure~\ref{plot-0142} to show the data
before and after this latest glitch, making the low-level changes in
frequency due to timing noise hard to see. However, their presence is
clear in Panel~d of Figure~\ref{plot-res}. 
%% The large 2011 glitch was accompanied by marginal
%% ($2\sigma$ level) changes in the pulsed flux in all
%% analysed bands (Panel~e of Figure~\ref{plot-0142}).

%% --------------------------------------------------------
%% --------------------------------------------------------
%% --------------------------------------------------------
%% --------------------------------------------------------

\subsection{1E~1048.1$-$5937}
\label{sec:1048}

\begin{figure}
\centerline{\includegraphics[scale=0.70]{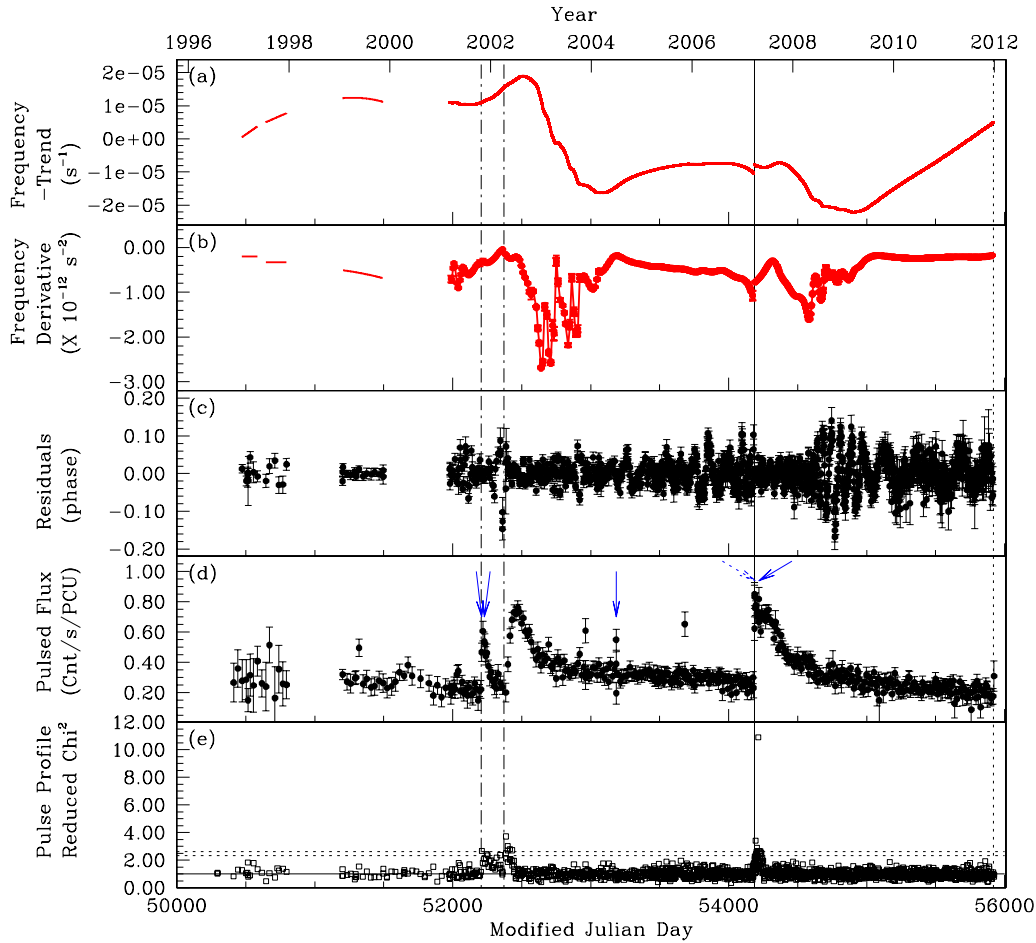}}
\caption
{ %% open caption
\footnotesize
Long-term evolution of the properties of 1E~1048.1$-$5937.
(a) Curves before the year 2000: ephemerides from \cite{kgc+01}.
Curve after the year 2000: frequency of 1E~1048.1$-$5937 as a function of
time with a linear trend subtracted, obtained from spline fitting.
%% Data points: the spline evaluated at the epoch of the observations.
(b) Curves before the year 2000: ephemerides from \cite{kgc+01}.
Curve after the year 2000:
frequency derivative as a function of time, obtained from
spline fitting. Data points: the spline evaluated at the epoch of
the observations. 
(c) Timing residuals obtained after subtracting
the TOAs from the ephemerides plotted in Panels~a and~b.
(d) RMS pulsed flux in the 2$-$20~keV band binned every 7 days.
The solid arrows mark the locations of the {\emph{RXTE}}-detected bursts.
The dotted arrow marks the location of a possible {\emph{RXTE}}-detected burst.
(e) Reduced ${\chi}^{2}$ statistics for the pulse profiles versus
time, calculated after
subtracting the scaled and aligned profiles of the individual
observations from a high signal-to-noise template for 1E 1048.1$-$5937.
The solid horizontal line indicates a reduced ${\chi}^{2}$ of 1.
The lower dotted line corresponds to the 2$\sigma$ significance level,
after having taken the number of trials into account.  
The upper dotted line corresponds to the
3$\sigma$ significance level.
%% dot dash
The dot-dashed vertical lines indicate the onset of the first two
pulsed flux flares. 
The solid vertical line indicates the location of a glitch (see Table~\ref{TBIG}). 
The dotted line shown in 2012 is placed just before the last
set of 3 {\emph{RXTE}} observations of 1E~1048.1$-$5937, which were anomalous (see \S~\ref{sec:1048}).
\normalsize
\label{plot-1048}
} %% close caption
\end{figure}

\begin{figure}
\centerline{\includegraphics[scale=0.70]{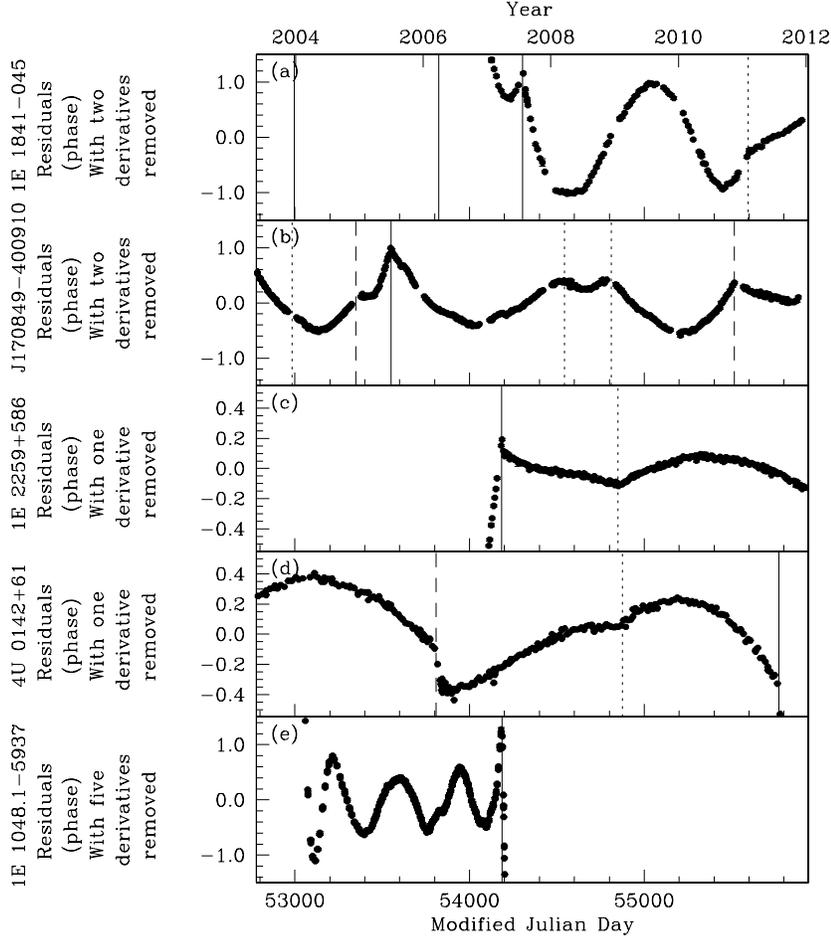}}
\caption
{ %% open caption
%% \footnotesize
Timing residuals of the five AXPs for selected stretches of time. The
residuals are obtained after subtracting the measured TOAs
from those predicted by an ephemeris consisting of
a frequency and of a small number of frequency derivatives.
The number of derivatives was chosen in such a way as to avoid
flattening the residuals, in order to show the changes in the curvature
of the residuals caused by timing discontinuities.
(a) Timing residuals for 1E~1841$-$045 
corresponding to an ephemeris
consisting of a $\nu$,
$\dot{\nu}$, and $\ddot{\nu}$ (a quadratic trend in frequency).
(b) Timing residuals for RXS~J170849.0$-$400910
corresponding to an ephemeris
consisting of a quadratic trend in frequency.
(c) Timing residuals for 1E~2259+586
corresponding to an ephemeris consisting of a linear trend in frequency.
(d) Timing residuals for 4U~0142+61
corresponding to an ephemeris consisting of a linear trend in frequency.
(e) Timing residuals for 1E~1048.1$-$5937
for the period of time when the AXP shows the least timing noise.
The residuals correspond to an ephemeris
consisting of the frequency $\nu$ and four frequency derivatives.
Solid lines indicate glitches,
dashed lines indicate glitch candidates,
and dotted lines indicate other timing discontinuities.
%% \normalsize
\label{plot-res}
} %% close caption
\end{figure}

\begin{figure}
\centerline{\includegraphics[scale=0.70]{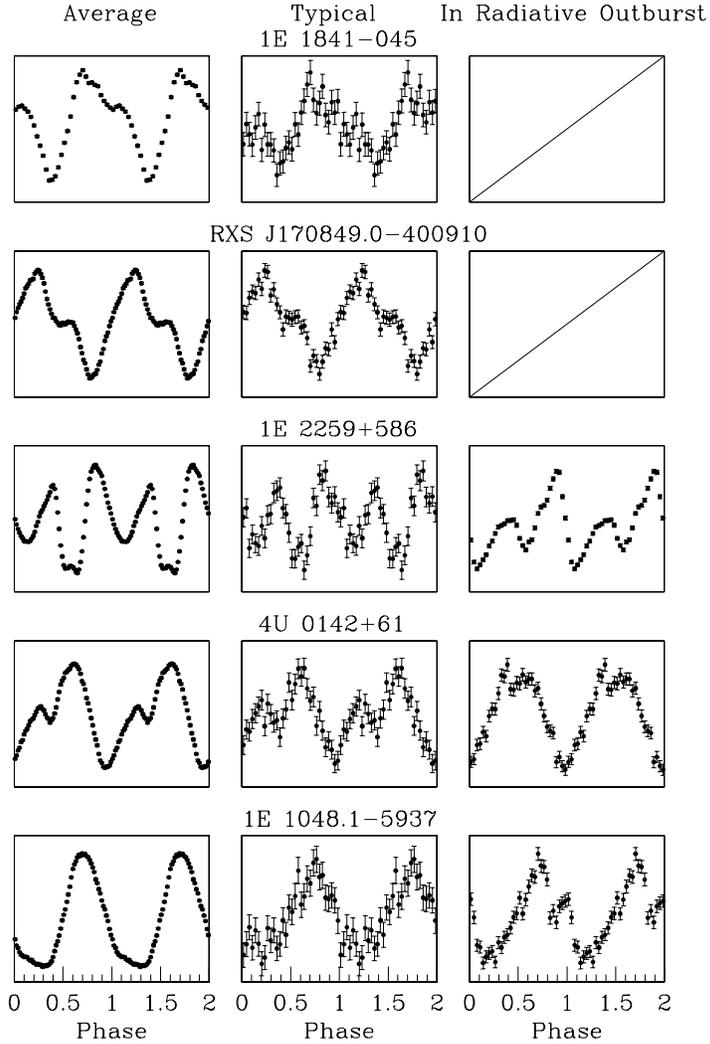}}
\caption
{ %% open caption
%% \footnotesize
%% The pulse profiles of the five monitored AXPs.
%%
Pulse profiles from each of the five monitored AXPs in the following
energy bands: 
1E~1841$-$045: 2.0$-$11.0~keV,
1RXS~J170849.0$-$400910: 2.0$-$5.5~keV,
1E~2259+586: 2.5$-$9.0~keV,
4U~0142+61: 2.5$-$9.0~keV,
and 1E~1048.1$-$5937: 2.0$-$10.0~keV.
{\emph{Left-Most Column:}} the long-term average
pulse profile. {\emph{Second Column:}} pulse profiles from
five typical observations, to show our usual data quality.
%% %% the deviation from the long-term average
%% %% profile is $\sim$~1~$\sigma$ in all five cases.
%% %% {\emph{Third Column:}} five unusual
%% %% pulse profiles obtained outside periods of outburst.  the deviations
%% %% from the long-term average profiles are, from top to bottom,
%% %% 2.1~$\sigma$, 3.4~$\sigma$, 3.1~$\sigma$, 1.7~$\sigma$,
%% %% and~2.9~$\sigma$. 
{\emph{Right-Most Column:}} pulse profiles of three
AXPs in outburst. 
%% %% the deviations from the long-term average profiles
%% %% are, from top to bottom, 101.2~$\sigma$, 8.1~$\sigma$,
%% %% and~5.3~$\sigma$. These numbers are old.. From the thesis..
AXPs 1RXS~J170849.0$-$400910 and 1E~1841$-$045 did not go into a
radiative outburst while being monitored by {\emph{RXTE}}. Note that
the various plots in this Figure have different net integration times.
%%
%% \normalsize
\label{plot-profiles}
} %% close caption
\end{figure}

A summary of the behavior of 1E~1048.1$-$5937 between 1997 and 2012 as
seen by {\emph{RXTE}} is shown in Figure~\ref{plot-1048}. Long-term
spin parameters are not presented in a Table like those of the other
AXPs because of the very large timing noise of the source,
necessitating multiple short-term ephemerides with a large number of
frequency derivatives.

To visually appreciate the strength the timing noise of the source,
refer to Panel~e of Figure~\ref{plot-res}. The timing residuals for
the period of time during which the pulsar was exhibiting the
{\emph{least}} timing noise are presented, after the subtraction of an
ephemeris containing five frequency derivatives.  Similar amplitude
residuals with other AXPs can generally be obtained with only one or
two frequency derivatives.

From a timing point of view, the period of time during which
1E~1048.1$-$5937 was observed by {\emph{RXTE}} was very eventful
(Panels~a and ~b of Figure~\ref{plot-1048}). First, from 1996 to 2001,
the large timing noise and the sparsity of the data made it possible
to only obtain phase-connected timing solutions for short (i.e.
months-long) periods of time \citep{kgc+01}. In 2001, we adopted the
strategy of observing the source in sets of three closely spaced
observations, making phase connection easier.

The source exhibited two slow-rise pulsed flux flares (time scale
weeks to months) in 2001 and 2002 (Panel~d of Figure~\ref{plot-1048}),
previously reported by \cite{gk04} and \cite{gkw06}. The flares were
accompanied by variations in the frequency derivative (Panel~b of
Figure~\ref{plot-1048}), although not as large or rapid as the
dramatic changes of two orders of magnitude in the rotational
frequency derivative which occured approximately a year later,
starting in 2001 November (Panel~b of Figure~\ref{plot-1048}). The
period of dramatic frequency derivative changes lasted approximately
450 days, and was followed by a period of radiative quiescence during
which there was significantly less timing noise (though still 
significantly more than in the other AXPs, see Figure~\ref{plot-res}).

On 2007 March~26, {\emph{RXTE}} detected a large glitch from the source
(see Table~\ref{TBIG}), and the pulsed flux started rising again
(Panel~d of Figure~\ref{plot-1048}). Once again, less than a year
after the pulsed flux started rising, another episode of large and
rapid variations in the frequency derivative were observed.  
%% The repeated pattern may suggest cyclic behavior.

Short bursts were observed on at least 4 occasions by {\emph{RXTE}}
(marked by arrows in Panel~d of Figure~\ref{plot-1048}), and on one
other occasion where a large burst occured too close to the end of the
observation, making it impossible to verify follow-up fluctuations in
the pulsed flux within that observation. Deviations from the
sinusoidal-looking pulse profile (see Figure~\ref{plot-profiles})
accompanied all three rises in the pulsed flux.

The last set of three consecutive {\it RXTE} observations of this
source, taken on 2011 December~28, were anomalous: timing residuals
were offset 0.15 in phase relative to the local ephemeris, and there
was a hint of an increase in the pulsed flux. 
%% However, given the tremendous timing noise of
%% this source, more data are needed to confirm whether an event
%% occured at this epoch.  
Fortunately subsequent {\it Swift} observations were made and in fact
did confirm this was the start of a new event; these will be reported
on elsewhere (Archibald et al. in prep.).

%% --------------------------------------------------------
%% --------------------------------------------------------
%% --------------------------------------------------------
%% --------------------------------------------------------
%% \clearpage

\subsection{Glitch Table}
\label{sec:tablenotes}

This section contains the notes for Table~\ref{TBIG}.

%% \footnotesize
%% \subsubsubsection{Table Notes}
%% a.
%% \normalsize
%%
%% \footnotesize

\begin{itemize}

\item a. 
When the TOAs near a timing event can be fitted by a sudden jump in
frequency as well as by an ephemeris consisting of 4 or 5 frequency
derivatives, the event is labelled a ``glitch candidate.'' When the
TOAs near a timing event can be fitted by a sudden jump in frequency
as well as by an ephemeris consisting of $\leq$3 frequency
derivatives, the event is labelled a ``notable timing discontinuity.''
Such discontinuities are more common, and only reported in this Table
when they are associated with radiative changes.

\item b. 
MJD range used for fitting the glitch.  We fit for a single $\nu$ and
$\dot{\nu}$ before the glitch and a sudden jump in $\nu$ and
$\dot{\nu}$ at the time of the glitch. It is important to note that
the value of the jump in $\dot{\nu}$ is very dependent on the amount
of data used. We limited the period of time fitted, but included
enough data to constrain $\dot{\nu}$.

\item c. 
Total frequency jump at the glitch epoch.

\item d. 
Total fractional frequency jump at the glitch epoch.

\item e. 
For the glitches where no recovery was observed immediately after the
glitch, this parameter represents the total jump in the frequency
derivative at the glitch epoch. For the glitches with an exponential
recovery, this represents the long-term change in the frequency
derivative. This parameter is extremely sensitive to the amount of
data included when doing the glitch fit, especially for noisy sources.

\item f. 
Fraction $Q$ of total $\Delta\nu$ recovered, and timescale of the
exponential recovery, if any.

\item g. 
Known radiative events associated with the glitch and observed with
{\emph{RXTE}}.

%% \item g. 
%% Estimate of the long-term value of $\dot{\nu}$ from 12 years
%% of {\emph{RXTE}} monitoring.

%% \item h. 
%% Magnetic fields here are calculated via $B \equiv 3.2 \times 10^{19}
%% \sqrt{P \dot{P}}$~G, where $P$ is the spin period in seconds and $\dot{P}$
%% is the period derivative.

\item h. 
\cite{dkg08} reported two possible sets of parameters for the first
glitch from 1E~1841$-$045. Subsequently, with the help of two
additional TOAs extracted from {\emph{XMM}} data, we were able to
constrain the fit parameters and show that the timing solution that we
reported as least likely for that glitch (the one with the smaller
frequency jump) was the correct one \citep{RIMASPEN, RIMTHESIS}.

\item i. 
See \S\ref{sec:1841} for details.

%% \item j. 
%% Entries with the value `0' are consistent with being zero.

%% ((degree))
\item j. 
These two candidate glitches occurred in or near an observing gap. The
frequency after the gap was higher than expected given the pre-gap
ephemeris. Because the pre-gap and post-gap combined TOAs can be fit
easily with a polynomial of degree 4 or 5, it is classified as a
glitch candidate \citep{dkg08}.

\item k. 
\cite{igz+07} classified this event as a glitch.

\item l. 
The difference between the value reported in \cite{igz+07} and that
reported in \cite{dkg08} can be attributed to the number of TOAs used
when fitting for the glitch parameters.

\item m. 
The difference between the sign of the value reported in \cite{igz+07}
and that reported in \cite{dkg08} can be attributed to the number of
TOAs used when fitting for the glitch parameters.

%% WAS THERE FOR CONSISTENCY, BUT ELIMINATED BECAUSE NOT NECESSARY
%% \item o. In \cite{dkg08}, we reported on three candidate glitches from
%% RXS~J170849.0$-$400910. Only the second of those appears in this
%% table. The first event was smaller in magnitude, and the source
%% exhibits such events often, so this event is now marked as a timing
%% discontinuity. The third event was also smaller in magnitude, and was
%% too close at the time to the end of the data set, which increased the
%% uncertainties in the timing parameters.
%X X
%% \item p. This new glitch candidate occured near an observing gap.

\item n. 
The model used to fit this glitch consists of a combination of rising
and falling exponentials; see \cite{wkt+04} for details. The
$\Delta\nu$ reported here is the maximum $\Delta\nu$ observed when
comparing the pre-glitch and post-glitch frequencies.

%% \item p. 
%% Entries with the value `0' are consistent with being zero.

\item o. 
See \S\ref{sec:2259} for details.

\item p. 
Panel~a of Figure~\ref{plot-0142} shows that a glitch might have
occurred in the observing gap. The ranges of $\Delta\nu$ and of
$\Delta\dot{\nu}$ reported here were obtained by extending the pre-gap
ephemeris (consisting of a $\nu$ and a $\dot{\nu}$) forward, and the
best-fit ephemeris of the two years after the gap (consisting of a
frequency and a frequency derivative) backward to the gap boundaries.

\item q. 
\cite{mks05} confined the glitch epoch to the 50893$-$51390 range of
MJDs, where 50893 is the MJD for the last pre-gap {\emph{RXTE}}
observation of 4U~0142+61, and 51390 is the first inter-gap
{\emph{ASCA}} observation of the source.

\item r. 
This is classified as a candidate glitch because the claim of a large
sudden frequency jump is based on a single observation, the first of
an active phase which the source entered in 2006. The TOA for that
observation might be affected by pulse profile changes \citep{gdk11}.
If this TOA is omitted, the initial sudden spin-up is less
significant, although it is clear that a change in $\dot{\nu}$
occured.  Also, even if this TOA is omitted, extending the
post-recovery ephemeris backward in time makes it look as though an
`anti-glitch' occurred \citep{gdk11}.  See \S\ref{sec:0142} for details.

\item s. 
The glitch marked the onset of an active phase in which there was a
short-term (within individual observations) pulsed flux increase
associated with the bursts \citep{gdk11}. There also was a subtle
29$\pm$8\% increase in the pulsed flux in the 2$-$10~keV band in the
years preceding the active phase which might have been associated with
the glitch \citep{dkg07,gdk+10}.
%% 52400 TO 53350  29 pm 8

\item t. 
See \S\ref{sec:0142} for details.

%% \item w. 
%% Starting in 2002 November, and for the next 450 days, AXP
%% 1E~1048.1$-$5937 exhibited extreme $\dot{\nu}$ noise
%% \citep{gk04,dkg09}. Several glitches might have occurred during this
%% time period. The glitches that we are reporting here are from outside
%% this period. There also has been unusual noise and a timing anomaly
%% near the onset of the first pulsed flux flare where a glitch might
%% have occurred. In the table, we are reporting on the glitches that
%% accompanied the second and the third pulsed flux flares from this
%% source.

\item u. 
The timing noise in this source is always large compared to that of
the other AXPs, most notably starting in 2002 November for a period of
450 days. Large variations in the rotational frequency derivative on
timescales of weeks to months occured multiple times throughout the
monitoring program, including ones that coincided with the onset of
the two pulsed flux flares. We choose not to report these rapid
changes as glitches as there would be too many.  The timing noise was
also particularly large at the onset of the two pulsed flux flares in
late 2001 and early 2002, both of which were accompanied by timing
anomalies of uncertain nature \citep{dkg09}. In this Table however, we
are only reporting on the large glitch that occured in 2007, marking
the end of a particularly quiet period (in timing and radiatively).

%% \item x.
%% The
%% parameters of this glitch are not very well constrained because of
%% unusually large pulse profile changes at the onset of the flare that
%% is associated with the glitch \citep{dkg09}. This glitch is classified
%% as a candidate glitch because the source was so noisy near this epoch,
%% it is hard to distinguish rapid noise from sudden jumps in the
%% frequency and frequency derivative.

\item v. 
There was a hint of a radiative and timing anomaly in the last set of
three {\emph{RXTE}} observations of 1E~1048.1$-$5937. However, the
lack of follow-up {\it RXTE} monitoring observations made it
impossible to determine whether a glitch, marking the end of another
quiet period, occured at that time.  Subsequent {\it Swift}
observations confirm it was the start of a new event, and will be
reported on elsewhere (Archibald et al., in prep.).

\end{itemize}

\normalsize

%% --------------------------------------------------------
%% --------------------------------------------------------
%% --------------------------------------------------------
%% --------------------------------------------------------

\section{Discussion}
\label{sec:discussion}

We have reported on roughly 16~years and 10~Ms of regular, systematic
{\it RXTE} monitoring of five AXPs on a weekly to monthly basis.  This
has yielded a rich and unique data set with which to examine magnetar
rotational and radiative behavior.  In particular, these data allow us
to try to identify common behavior and interesting differences from
source to source.  Indeed, what is clear is that each source has its
own `character', with no two of our targets behaving very similarly.
For example, apart from well defined and highly localized timing
anomalies, 1E~2259+586 is a very stable rotator, which contrasts
strongly with 1E~1048.1$-$5937, which, even at its most stable, is
still wildly erratic by comparison. That said, a systematic
consideration of the data sets reveals some clear common behavior
that can hopefully serve as a useful resource for developing a
coherent theory to explain magnetar phenomenology.

Specifically, there are clear patterns in the radiative and timing
anomalies in our targets. In Table~\ref{table-summary}, we present a
concise summary of the different types of behaviors we have observed,
in particular a `scorecard' showing the fraction of observed anomalies
that are accompanied by a different type of interesting behavior. In
the Table, for each phenomenon listed in the far left column, we
indicate in its row the fraction of times it is accompanied by the
phenomenon listed at the top of each column.  For example, of a total
of 22 timing anomalies (including unambiguous glitches) observed in
our full data set, 5 were accompanied by some form of flux change.

%%%%%% Several points are clear from examination of this Table: 
%%%%%% Several trends are clear from examination of this Table:
Several points are clear from examination of this Table:

\noindent %%
%% 5 long lived flux var: 2259(2), 1048(3), 0142(0), 
%% bursting cases: 1841(once), 1048(flare1, flare2, glitch), 0142(active), 2259(major).
%% profile cases: 2259(2), 0142(1), 1048(3)
(i) Radiative changes in our target AXPs are rare.  In $\sim$16~yr of
monitoring per source, we have detected only 5 long-lived episodes of
flux variations, and only 6 cases of SGR-like bursting (where we
define the 2002 outburst of 1E 2259+586 as a single case of bursting,
in spite of there having been observed over 80 SGR-like bursts at one
epoch). One caveat regarding this observation is that our {\it RXTE}
monitoring was only sensitive to pulsed flux changes, whereas total
flux changes can sometimes be larger \citep[e.g.][]{tmt+05}. Our
observations set the approximate time scale for radiative outbursts in
AXPs at the several year level, at least for our sample. The latter is
biased toward those sources that are brightest in quiesence; this
inferred time scale may not apply to the so-called transient magnetars
\citep[e.g.][]{hg05}. Also note we find evidence for constant
low-level pulse profile and flux changes generically as discussed in
\S\ref{sec:results}, so though large radiative changes appear rare, at
lower levels, they seem common.

\noindent %%
(ii) Radiative changes are almost always accompanied by some form of
timing anomaly.  In all cases, flux enhancements and/or pulse profile
changes were accompanied by some form of glitch or timing anomaly, and
5/6 burst episodes were similarly accompanied.

\noindent %%
(iii) The converse of (ii) is not true: only occasionally (20-30\% of
the time) are timing anomalies in AXPs accompanied by any form of
radiative anomaly.  Equivalently, the large majority of AXP timing
anomalies are radiatively silent, 
%% sentence below added by Rim
particularly but not uniquely in RXS~J170849.0$-$400910 and 1E~1841$-$045, the two AXPs
that glitch most often.
%% sentence above added by Rim
This implicates the silent glitches as having an origin in the stellar
interior, because if the rotational anomaly were due to physical
changes in the magnetosphere, via enhanced currents for example, it
seems likely that profile and/or flux changes should be present as
well.  Moreover magnetospheric current changes seem unlikely to
produce abrupt spin-up events so similar to those seen in radio
pulsars having far lower B fields \citep[e.g.][]{elk+11,ymh+13}.
%% no ref for this last bit. Now yes!

%% below: largest glitch yet seen where???
Further, it is not necessarily the largest timing anomalies (i.e. the
glitches with largest fractional frequency change) that are
accompanied by radiative changes.  Indeed using our strictest glitch
definition, only 2 of 11 certain glitches had radiative changes.
However, it seems possible that the greater the timing change, the
greater the probability of there being a radiative change; for
example, the largest glitch yet seen (see Table~\ref{TBIG}), in
1E~1048.1$-$5937, was accompanied by substantial radiative changes.
Moreover there is no evidence that any other glitch property, such as
recovery, is preferentially associated with radiative changes.
Furthermore, the larger radiative outbursts do not necessarily
correspond to the larger timing anomalies; for example the first of
the two large flares in 1E~1048.1$-$5937 had only a small timing
anomaly.

Note that there is significant overlap in properties of AXP glitches
($\Delta\nu$, $\Delta\dot{\nu}$) that do and do not have accompanying
radiative behavior, and there are no particular timing differences
between silent and loud timing events. These facts, together with the
absence of obvious mechanisms to produce sudden radiatvely silent
spin-ups in the magnetosphere, argue in favor of the hypothesis that
{\it all} AXP glitches have their physical origin in the stellar
interior, as has long been hypothesized in radio pulsars
\citep[e.g.][]{ai75,pa85}. Why some interior events produce radiative
output while others do not could be related, for example, to the
amount of energy released internally, or to the depth at which the
event occurred.

\noindent %%
(iv) When one type of radiative change occurs, usually multiple types
of radiative changes occur.  For example, pulse profile changes
are, the large majority of the time, accompanied by a flux increase
and SGR-like bursts.
This clearly implicates external changes, though in our view these
must accompany the internal events that caused the accompanying timing anomaly.

%% Kes75, 1119.
We note additionally that the apparently magnetar-like events observed
in an otherwise conventional rotation-powered pulsar also conformed to
the above trends.  Specifically, in the 2006 magnetar-like outburst
of the 0.326-s rotation-powered pulsar PSR~J1846$-$0258
\citep{ggg+08}, the pulsar suffered a large rotational glitch in
addition to SGR-like bursts and a large X-ray flux increase (but no
apparent pulse profile changes).  In this way, this event agreed with
pattern (i), since such events are rare in this source, having
occurred just once between 1999 and 2013, with (ii) because the
radiative changes were accompanied by a timing anomaly, with (iii)
because previously the source had glitched and showed no radiative
changes \citep{lkgk06}, and with (iv) because both a flux change and
SGR-like bursts occurred. Interestingly the only radiative anomalies
yet to have been observed in the high-B radio pulsar PSR J1119$-$6127
were also at the time of a glitch \citep{wje11}.  All
other radio pulsar glitches, were, as far as could be observed,
radiatively silent, although prompt target-of-opportunity X-ray
observations following glitches have not in general been done.
The lone exception was for the 1999 spin-up glitch of the Vela pulsar,
which was followed within a day  by a {\it Chandra} X-ray observation
that showed no change in flux \citep{hgh01}.

%% discovery paragraph
The majority of the magnetars presently known were discovered via
SGR-like bursts, thanks to the existence of sensitive all-sky X-ray
monitors such as the {\it Fermi} Gamma-Ray Burst Monitor and {\it
Swift}'s Burst Alert Telescope \citep[see][and references
therein]{ok13}.  These sources, discovered in outburst, were not being
monitored prior to discovery, so it is not known whether any form of
timing anomaly accompanied the outburst. However, in some cases there
are hints of glitch recovery post-outburst \citep[e.g.][]{snl+12},
suggesting a glitch did occur. 
The results of our {\it RXTE} monitoring program suggest that all or
nearly all magnetar outbursts are accompanied by timing anomalies;
careful monitoring of sources in quiescence in case of future
outbursts may be able to test this hypothesis.

%% physical properties
Do the behaviors of the five sources we have monitored correlate with
any of their inferred physical properties?  There are hints of this,
although with only five sources, all conclusions are tentative.
Nevertheless we note that the source that is least stable and most
prone to large radiative variations, 1E~1048.1$-$5937, is also the one
with the highest quiescent blackbody temperature (0.56$\pm$0.01 keV),
whereas the two most stable sources from a rotational point of view,
1E~2259+586 and 4U~0142+61, have the lowest quiescent temperatures
(0.400$\pm$0.007 and 0.410$^{+0.004}_{-0.002}$ keV, respectively),
with the remaining two sources, 1RXS~J1708$-$4009 and 1E~1841$-$045
having intermediate temperatures (0.456$\pm$0.009 keV and
0.45$\pm$0.03 keV, respectively), with all values as compiled by
\citet{ok13} and based on spectral fitting using the phenomenological
blackbody plus power-law models.  
%% B field stuff below by Rim.
We further note that 1E~2259+586 and 4U~0142+61, the two sources with
the least amount of noise in their spin-down trend, have the lowest
values of spin-inferred magnetic field strength (see
Table~\ref{table-sources}), and 1RXS~J1708$-$4009 and 1E~1841$-$045,
the two sources that glitch most frequently, have the largest values.
Thus, overall, our results suggest that higher-B sources and/or hotter
sources show more timing activity.  This is qualitatively as expected
in the magnetar model since higher B implies stronger field decay,
hence higher internal heating, greater interior stresses, and greater
likelihood of crust cracking and magnetic structure changes
\citep{tlk02,bel09,pp11a,pp11b}.

%% activity parameters
\citet{dkg08} calculated glitch activity parameters for the three AXP
targets that had glitched in $\sim$8-9~yr of {\it RXTE} monitoring.
They defined two different activity parameters: $a_g$ was the sum of
the initial fractional frequency changes (including decaying
components) divided by total observing span, and $A_g$ was the same
sum but in absolute value of the glitch (i.e. not fractional) again
averaged over the span.  We have recalculated these parameters for the
same three sources as well as for the other two (both of which have
now glitched) over the full {\it RXTE} span, where we take the
absolute value of of the frequency changes for anti-glitches.
Interestingly, we find a relatively narrow range of values for the
five sources:
$1.0 \times 10^{-14} < a_g < 3.3 \times 10^{-14}$~s$^{-1}$, and
$1.5 \times 10^{-15} < A_g < 5.1 \times 10^{-15}$~Hz~s$^{-1}$.
Here the uncertainties are no larger than $\sim$10\% of those values.
This suggests that the magnetar glitch rate is roughly constant among
sources, at least for the persistently bright AXPs that we have been
monitoring. We note that the largest values of $a_g$ and $A_g$ in the
above ranges come from 1E~1048.1$-$5937, 1E~1841$-$045 and RXS
J1708$-$4009, the three with the largest inferred B values.  This is
consistent with our inference above that timing activity is correlated
with that parameter.  Comparing with activity parameters for radio
pulsars compiled by \citet{dkg08}, the above range of AXP values of
$a_g$ are comparable to those of the most actively glitching radio
pulsars known, whereas the values of $A_g$ are among those of the
nearly most actively glitching pulsars known, but not quite as high as
those of some sources.  \citet{dkg08} also published histograms of
glitch amplitude distributions for AXPs and radio pulsars; updating
those results makes no qualitative difference to those histograms.

%% About to talk individually about 2259 and 1048.
Apart from the above overall trends, several other behaviors we have
observed are notable.

%% anti glitches
\citet{akn+13} recently reported a negative frequency step coincident
with an X-ray outburst from 1E 2259+586 as observed using the {\it
Swift} X-ray Telescope. They noted some previous possible
anti-glitches, including one previously mentioned by \citet{ibi12} for
this source. We have classified that event using our criteria as a
timing anomaly; however as our pulsed flux analysis shows, the event
was in fact coincident with a small X-ray flux enhancement (see Panel~d 
in Fig.~\ref{plot-2259}), as well as a brief, simultaneous change
in the pulse profile (Panel~e of Fig.~\ref{plot-2259}). 
%% sentence below changed by Rim because she thought it was unclear before.
These radiative changes greatly support the hypothesis that an anti-glitch 
occured in early 2009 in this source given the \citet{akn+13} results, 
%% end of change by Rim
although we note that the magnitude of the spin-down in this earlier
event is a factor of 4$-$9 smaller than in the {\it Swift} case, while
the X-ray pulsed flux increased by an amount similar to that in the
later event.
%% switch to 0142 anti
The possible anti-glitch we report in 4U~0142+61 in early 2009 was
not, however accompanied by any radiative changes, so we cannot
confirm its veracity in that way. 
%% back to 2259 anti
For 1E~2259+586, combining the {\it RXTE} and {\it Swift} events, the
source has thus exhibited 2 spin-up and 2 spin-down glitches in 17~yr
of monitoring; the equality of those numbers is striking, especially
given the absence of any evidence for anti-glitches in 3 other
AXPs we have monitored for comparable lengths of time, in spite of the
detection of many spin-up events. Overall our results suggest that
spin-down glitches in magnetars are much rarer than the spin-up
variety.
This is of course also true of radio pulsars, for which no spin-down
glitches have been reported in spite of hundreds of spin-up glitches
having been observed \citep[e.g.][]{elk+11,ymh+13}.

Additionally, in 1E 1048.1$-$5937, we have observed particularly
unusual variations in spin-down rate in two events that are localized
in time and both following relaxations from substantial pulse flux
enhancements.  Specifically, as previously reported by \cite{gk04}
and \citet{dkg09}, order-of-magnitude torque variations lasting
approximately one year were observed during the relaxations following
the 2002 and 2007 flux flares, both during times of relative flux
stability. The coincidental delay of the torque variations following
the flares is suggestive of a causal relationship, however the lack of
simultaneity of the two phenomena argue against accretion torque
variations, since that predicts no such delay. \citet{bel09} discusses
the possibility of delayed torque changes following a radiative
outburst in magnetars, and suggests this is due to the delay in
propagation of a disturbance in the field structure to the field-line
bundle nearest the magnetic pole, where the torque originates. Why a
single radiative outburst should result in the multiple torque
variations seen is unclear in his picture however. We note that monitoring of this
source using the {\it Swift} telescope, conducted after our {\it RXTE}
observations ceased, has detected a third such flux flare as well as
a subsequent episode of torque variations. This is discussed in
greater detail in an upcoming paper (Archibald et al., in prep.).

%% Or mention the swift monitoring here. 
%% this was my old summary paragrpah.
%% [[NOT SURE THIS PARAGRAPH IS STAYING]]. To sum up, Anomalous X-ray
%% Pulsars (AXPs) are unusual and active astronomical X-ray sources with
%% extreme properties. Much theoretical and observational progress has
%% been made since AXPs were recognized as a distinct class of sources.
%% On the theoretical front, the magnetar model was put forth to attempt
%% to explain both the properties of SGRs and those of AXPs, properties
%% which have been and are still studied with a variety of instruments
%% from the radio to the hard X-ray bands. 
%% The monitoring program of AXPs
%% with {\emph{the Rossi X-ray Timing Explorer}}, the findings of which
%% are summarized in this manuscript, has provided insight into the
%% timing properties of these objects, and by showing the various kinds
%% of radiative outbursts that these sources undergo. Indeed the
%% monitoring observations, combined with observations by imaging
%% instruments and observations outside the X-ray band, show that AXPs
%% exhibit a tremendous range of variable behavior, which poses a
%% challenge for any model that attempts to explain them. Thus, even
%% though they have been now studied for about two decades, there remains
%% much to be discovered and much to be explained about AXPs.
%% these mysterious objects.

\section{Conclusions}

The long-term, systematic monitoring program of AXPs by {\it RXTE} has
been valuable for illucidating the phenomenology of these unusual
objects.  With weekly to monthly data obtained over $\sim$16~yr for
five objects, we have discovered a variety of phenomena, including
outbursts and glitching, most of which strongly support the magnetar
model.  Here we have reported on repeated and remarkable large torque
variations in 1E~1048.1$-$5937, curiously occuring months after large
flux outbursts.  We also report on a small pulsed flux and pulsed
profile change at the time of a likely anti-glitch in 1E~2259+586.  We
have seen no radiative changes from 1RXS~J1708$-$4009, and only
apparent SGR-like bursts from 1E~1841$-$045
as far as radiative changes are concerned.
4U~0142+61 has shown a variety of low-level behaviors.  All our
targets have exhibited spin-up glitches at least once, and show
differing levels of timing noise in inter-glitch intervals.

In this overall compilation of all the {\it RXTE} AXP monitoring data,
we have noted the following patterns in AXP behavior: (i) large
radiative changes of any kind in our sample are rare, occurring at
most every several years; (ii) radiative changes in our AXPs are
almost always accompanied by some form of timing anomaly, usually a
spin-up glitch; (iii) the converse of (ii) is not true: only 20--30\%
of AXP timing anomalies are accompanied by any form of radiative
change, with some evidence that the likelihood of such changes
increases with glitch size; and (iv) when one type of radiative change
occurs (e.g. flux outburst) there are usually others (e.g. pulse
profile changes or bursts).  There is an apparent tendency for the
sources with higher spin-inferred magnetic fields and higher measured
surface temperatures to be the most timing active, although the
paucity of monitored sources makes this conclusion weak. The estimated
glitch activities of all our targets are within a factor of $\sim$3--5
of each other (depending on precise definition), and are similar to
those of the most actively glitching radio pulsars.  Overall, the
overlap in timing properties involved in spin-up and spin-down
glitches seen thus far, together with the overlap of AXP glitch
behavior compared with that seen in the radio pulsar population, is
consistent with {\it all} glitching having origin in the stellar
interior, and may be hinting at structural differences between high-
and low-B neutron stars.  On the other hand, the stellar magnetosphere
may also play a role in the timing anomalies, especially in the
radiative changes that sometimes accompany them.\footnote{Very
recently, just prior to the submission of this paper, Lyutikov (2013;
arXiv1306.2264) and Tong (2013; arXiv1306.2445) suggested that
external mechanisms could explain the 1E~2259+586 anti-glitch reported
by Archibald et al. (2013). However as we have argued here, we believe
the systematic observations reported on here challenge those
interpretations.}

%% which subtle differences are referred to below???
Although greatly illuminating of AXP behavior, these {\it RXTE}
observations have raised many interesting questions that remain
unanswered.  Particularly noteworthy is the great diversity in our
targets' behavior, which ranges from near-complete radiative stability
over the 16-yr interval (1RXS J1708$-$4009 and 1E 1841$-$045) to large
flux changes (1E~2259+586 and 1E~1048.1$-$5937).  Moreover, the
difference in timing stability from source to source is striking.
Further, the subtle differences in glitch properties in AXPs compared
to in radio pulsars is intriguing, and may be hinting at interesting
structural differences.  What role the magnetosphere may play in
timing events remains to be seen.  Continued systematic, long-term
monitoring of these and other sources is a powerful way to help
understand these objects, by better determining the statistics of
glitches and radiative outbursts, and hopefully correlating them with
physical properties of the sources in order to further constrain the
physics of these remarkable objects.

%% Vicky's acknowledgments, after Evan Smith and Divya Pereira fix. 
We are grateful for the support we received over the years from the
{\it RXTE} Guest Observers Facility, specifically Evan Smith and Divya
Pereira for assistance with scheduling and particularly Jean Swank 
for her support of this project over so many years.  We also thank Ed
Morgan and Kenton Phillips for kindly permitting us and often
assisting us in using MIT astrophysics computers and software for many
years. We also acknowledge the efforts of many people who have been
involved in the {\it RXTE} monitoring project over the years: Deepto
Chakrabarty, Fotis Gavriil, Eric Gotthelf, Jessica Lu, Julia
Steinberger, and Pete Woods as well as Andrei Beloborodov, Andrew Cumming, 
Rob Duncan, Jeremy Heyl, Kostas Gourgouliatos, Maxim Lyutikov, 
Chris Thompson, and Dave Tsang for useful conversations at various
times.  VMK holds the Lorne Trottier Chair in Astrophysics and
Cosmology, a Canada Research Chair, and is the R. Howard Webster
Foundation Fellow of the Canadian Institute for Advanced Research
(CIFAR).  This work was supported in part by an NSERC Discovery Grant,
the Centre de Recherche Astrophysique de Qu\'ebec via FQRNT, a CIFAR
Fellowship, a Killam Research Fellowship, and the NSERC John C. Polanyi
Award.

%% --------------------------------------------------------
%% --------------------------------------------------------
%% --------------------------------------------------------
%% --------------------------------------------------------

%%%%%%%%%%%%%%%%%% \bibliographystyle{apj}
%%%%%%%%%%%%%%%%%% \bibliography{journals1,extrarefs3,modrefs,psrrefs,crossrefs}
%% \begin{thebibliography}{62}
%% \expandafter\ifx\csname natexlab\endcsname\relax\def\natexlab#1{#1}\fi
%% \bibitem[{Alpar {\ldots}  
%% \end{thebibliography}

%% --------------------------------------------------------
%% --------------------------------------------------------
%% --------------------------------------------------------
%% --------------------------------------------------------
%% FIGURES
%% --------------------------------------------------------
%% --------------------------------------------------------
%% --------------------------------------------------------
%% --------------------------------------------------------

\onecolumn

%% --------------------------------------------------------
%% --------------------------------------------------------
%% --------------------------------------------------------
%% --------------------------------------------------------

%% >>>>>>>>>>>>>>>>>>>>>>>>>>>>>>>>>>>>>>>>>>>>>>>>>>  \ref{plot-1841}

\clearpage

%%% >>>>>>>>>>>>>>>>>>>>>>>>>>>>>>>>>>>>>>>>>>>>>>>>>>>>>\ref{plot-1708}

\clearpage

%%% >>>>>>>>>>>>>>>>>>>>>>>>>>>>>>>>>>>>>>>>>>>>> \ref{plot-2259}

\clearpage

%%% >>>>>>>>>>>>>>>>>>>>>>>>>>>>>>>>>>>>>>>>>>>>>>>> \ref{plot-0142}

\clearpage

%%% >>>>>>>>>>>>>>>>>>>>>>>>>>>>>>>>>>>>>>>>>>>>>>>>>>>>>>> \ref{plot-1048}

\clearpage

% >>>>>>>>>>>>>>>>>>>>>>>>>>>>>>>>>>>>>>>>>>>> \ref{plot-res}

\clearpage

% >>>>>>>>>>>>>>>>>>>>>>>>>>>>>>>>>>>>>>>>>>>> \ref{plot-profiles}

\clearpage

%% --------------------------------------------------------
%% --------------------------------------------------------
%% --------------------------------------------------------
%% --------------------------------------------------------
%% TABLES
%% --------------------------------------------------------
%% --------------------------------------------------------
%% --------------------------------------------------------
%% --------------------------------------------------------

%% >>>>>>>>>>>>>>>>>>>>>>>>>>>>> \ref{table-fotis}

\clearpage
\begin{deluxetable}{lccccc}
\small
\tabletypesize{\footnotesize}
\tablewidth{530.0pt}
%% \rotate
\tablecaption
{
Spin Properties of the {\it RXTE}-Monitored AXPs\tablenotemark{a}
\label{table-sources}
}
\tablehead
{Parameter &
1E~1841$-$045 &
RXS~J170849.0$-$400910 &
1E~2259+586 &
4U~0142+61 &
1E~1048.1$-$5937}
\startdata
Period, $P$ (s) & 11.78 & 11.00 & 6.98 & 8.69 & 6.46 \\
Frequency, $\nu \equiv 1/P$ (Hz) & 0.085 & 0.091 & 0.143 & 0.115 & 0.155 \\
$\dot{\nu}$ ($10^{-13}$ Hz/s)\tablenotemark{b} & $-2.95$ & $-1.60$ & $-0.099$ & $-0.268$ & $-5.33$ \\
Magnetic Field\tablenotemark{c},   $\; B$ (G) & $7.0 \times 10^{14}$  & $4.7 \times 10^{14}$ & $0.59 \times 10^{14}$ & $1.3 \times 10^{14}$ & $3.8 \times 10^{14}$ \\
%% $N_H$\tablenotemark{a}                  & 2.54            & 1.48             & 1.01         & 0.96             & 0.97   \\
%% $kT$\tablenotemark{b}                   & 0.44            & 0.46             & 0.37         & 0.39             & 0.656  \\
%% $n_{\mathrm{BB}}$\tablenotemark{c}      & 3.098           & 1.692            & 1.923        & 12.978           & 0.6006 \\
%% $\Gamma$\tablenotemark{d}               & 2               & 2.83             & 3.75         & 3.62             & 3.31   \\
%% $n_{\mathrm{PL}}$\tablenotemark{e}      & .660            & 2.461            & 4.26         & 15.5             & 0.415  \\
%% $f$\tablenotemark{f}                    & 2.51            & 2.45             & 1.41         & 7.2              & 0.48   \\
%% $\mathcal{PF}$\tablenotemark{g} [band]  & 18.9\% [0.6--7] & 29\% [6--10]     & 23\% [2--10] &  9.6\% [2.5--10] & 76\% [2--10] \\
%% References\tablenotemark{h,i}           & 1, 2            & 3                & 4            & 5                & 6 \\
%%
\enddata
\tablenotetext{a} {For additional information on these sources see \citet{ok13} or the McGill Online Magnetar Catalog\\ at http://www.physics.mcgill.ca/$\sim$pulsar/magnetar/main.html.}
\tablenotetext{b} {Estimate of the long-term value of $\dot{\nu}$ from
$\sim$16~yr of {\it RXTE} monitoring.}
\tablenotetext{c} {Magnetic fields calculated via $B \equiv 3.2 \times 10^{19} \sqrt{P \dot{P} }$~G.}
%% \tablenotetext{d} {Power-law index.}
%% \tablenotetext{e} {Normalization of the power-law component in units of photons keV$^{-1}$~cm$^{-2}$~s$^{-1}$ at 1~keV.}
%% \tablenotetext{f} {Unabsorbed flux in the 2--10~keV band in units of $10^{-11}$~erg~s$^{-1}$~cm$^{-2}$.}
%% \tablenotetext{g} {Pulsed fraction (\%) in the energy band (in keV) denoted in the square brackets.}
%% \tablenotetext{h} {References: (1)~\citet{mskk03}; (2)~\citet{ks10}; (3)~\citet{roz+05}; (4)~\citet{zkd+08}; (5)~\citet{gws05}; (6)~\citet{tgd+08}.}
%% \tablenotetext{i} {This table has also made use of information from the online magnetar catalog provided by the McGill University Pulsar Group
%% (\url{http://www.physics.mcgill.ca/~pulsar/magnetar/main.html})}
\end{deluxetable}

%% >>>>>>>>>>>>>>>>>>>>>>>>>>>>>>>>>>>> (the observations table) \ref{TOBS}

\clearpage
\begin{deluxetable}{llllll}
\tabletypesize{\scriptsize}
\tablewidth{420.0pt}
%% \rotate
\tablecaption
{
{Summary of \emph{RXTE}} data analyzed for target AXPs
\label{TOBS}
}
\tablehead
{
Source &
Number of &
Typical &
Date of the &
Date of the &
Band used for
%% &
%% $$ 
\\
&
analysed PCA &
exposure per &
first analysed &
last analysed &
the timing 
%% &
%% {\emph{RXTE}}-detected 
\\
&
observations &
observation &
observation &
observation &
analysis 
%% &
%% bursts
}
\startdata
%% %1E~1841$-$045 & 285 & 281\tablenotemark{a} & 5~ks & 08/31/1996 & 12/08/2011 & 2$-$11~keV & 1\\
%% %$$ & $$ & $$ & $$ & $$ & $$ & $$ & $$\\
1E~1841$-$045 &  281\tablenotemark{a} & 5~ks & 08/31/1996 & 12/08/2011 & 2$-$11~keV\\
%% %RXS~J170849.0$-$400910 & 528 & 522 & 2~ks & 01/12/1998 & 11/17/2011 & 2$-$5.5~keV & 0\\
%% %$$ & $$ & $$ & $$ & $$ & $$ & $$ & $$\\
RXS~J170849.0$-$400910 &  522 & 2~ks & 01/12/1998 & 11/17/2011 & 2$-$5.5~keV\\
%% %1E~2259+586 & 622 & 608 & 5~ks & 09/29/1996 & 12/29/2011 & 2.5$-$9~keV & $\sim$80,\\
%% %$$ & $$ & $$ & $$ & $$ & $$ & $$ & within 1 observation.\\
1E~2259+586 &  608 & 5~ks & 09/29/1996 & 12/29/2011 & 2.5$-$9~keV\\
%% %4U~0142+61 & 344 & 339 & 5~ks & 03/28/1996 & 12/22/2011 & 2.5$-$9~keV & 6,\\
%% %$$ & $$ & $$ & $$ & $$ & $$ & $$ & within 3 observations.\\
4U~0142+61 &  339 & 5~ks & 03/28/1996 & 12/22/2011 & 2.5$-$9~keV\\
%% %1E~1048.1$-$5937 & 1456 & 1452 & 2~ks & 07/29/1996 & 12/29/2011 & 2$-$5.5~keV & 4,\\
%% %$$ & $$ & $$ & $$ & $$ & $$ & $$ & within 4 observations.\\
1E~1048.1$-$5937 & 1452 & 2~ks & 07/29/1996 & 12/29/2011 & 2$-$5.5~keV\\
\enddata
\tablenotetext{a} {TOAs obtained from two {\emph{XMM}} observations
(2001 October~7 and 2002 October~5) were used
to supplement those extracted from {\emph{RXTE}} observations of
1E~1841$-$045; see \S\ref{sec:timinganalysis} and \ref{sec:1841} for details.}
\end{deluxetable}

% >>>>>>>>>>>>>>>>>>>>>>>>>>>>>>>>>>>>>>>>>>>>>>> \ref{table-1841}

\clearpage
\begin{deluxetable}{lccccccc}
%% \tabletypesize{\scriptsize}
\tabletypesize{\footnotesize}
\tablewidth{690.0pt}
\rotate
\tablecaption
{
Long-Term Spin Parameters for 1E~1841$-$045\tablenotemark{a}
\label{table-1841}
}
\tablehead
{
&
\colhead{Ephemeris A} &
\colhead{Ephemeris B1\tablenotemark{b,c}} &
\colhead{Ephemeris B2\tablenotemark{b}} &
\colhead{Ephemeris C} &
\colhead{Ephemeris D} &
\colhead{Ephemeris E} &
\colhead{Ephemeris F}\\
\colhead{Parameter} &
\colhead{Spanning MJD} &
\colhead{Spanning MJD} &
\colhead{Spanning MJD} &
\colhead{Spanning MJD} &
\colhead{Spanning MJD} &
\colhead{Spanning MJD} &
\colhead{Spanning MJD}\\
&
\colhead{51225$-$52438} &
\colhead{52482$-$52774} &
\colhead{52610$-$52981} &
\colhead{53030$-$53816} &
\colhead{53829$-$54280} &
\colhead{54308$-$55538} &
\colhead{55616$-$55903}
}
\startdata
%%                                           & A               & Be                 & Bl             & C              & D                & E              & F
MJD start                                    & 51224.538       & 52481.594          & 52610.313      & 53030.093      & 53828.808        & 54307.465      & 55615.911\\
MJD end                                      & 52437.712       & 52773.487          & 52981.186      & 53815.842      & 54279.628        & 55538.066      & 55903.345\\
TOAs                                         & 56              & 9\tablenotemark{b} & 17             & 55             & 32               & 91             & 21\\
$\nu$ (Hz)                                   & 0.0849253010(8) & 0.0848981884(13)   & 0.084888570(3) & 0.084889876(6) & 0.084868826(4)   & 0.084857001(4) & 0.084824992(7)\\
$\dot{\nu}$ (10$^{-13}$ Hz s$^{-1}$)         & $-$2.9954(7)    & $-$3.158(6)        & $-$3.179(2)    & $-$3.340(7)    & $-$2.834(5)      & $-$2.866(6)    & $-$2.944(11)\\
$\ddot{\nu}$ (10$^{-22}$ Hz s$^{-2}$)        & 3.13(11)        & $-$10.1(9)         & $-$            & 16.0(4)        & $-$4.3(2)        & $-$3.2(4)      & $-$3.9(7)\\
$d^{3}\nu/dt^{3}$ (10$^{-29}$ Hz s$^{-3}$)   & 1.29(6)         & $-$                & $-$            & $-$2.78(11)    & $-$              & 2.86(17)       & $-$\\
$d^{4}\nu/dt^{4}$ (10$^{-36}$ Hz s$^{-4}$)   & $-$2.23(14)     & $-$                & $-$            & $-$            & $-$              & $-$0.80(3)     & $-$\\
$d^{5}\nu/dt^{5}$ (10$^{-44}$ Hz s$^{-5}$)   & 9.0(8)          & $-$                & $-$            & $-$            & $-$              & $-$            & $-$\\
Epoch (MJD)                                  & 51618.0001      & 52692.0000         & 53006.0000     & 53006.0000     & 53822.0000       & 54305.0000     & 55585.0000\\
RMS residual (phase)                         & 0.031           & 0.014              & 0.028          & 0.036          & 0.024            & 0.041          & 0.017\\
\enddata
\tablenotetext{a} {Numbers in parentheses are TEMPO-reported 1$\sigma$ uncertainties. } 
\tablenotetext{b} {The B ephemeris was split into two segments
because of the presence of a significant second frequency derivative for
the first few months following the 2002 glitch.}
\tablenotetext{c} {Including TOAs from two {\emph{XMM}} observations.}
\end{deluxetable}

% >>>>>>>>>>>>>>>>>>>>>>>>>>>>>>>>>>>>>>>>>>>>>>> \ref{table-1708}

\clearpage
\begin{deluxetable}{lccccccccc}
%% \tabletypesize{\scriptsize}
\tabletypesize{\tiny}
\tablewidth{675.0pt}
\rotate
\tablecaption
{
Long-Term Spin Parameters for RXS~J170849.0$-$400910\tablenotemark{a}
\label{table-1708}
}
\tablehead
{
&
\colhead{Ephemeris A} &
\colhead{Ephemeris B} &
\colhead{Ephemeris C} &
\colhead{Ephemeris D} &
\colhead{Ephemeris E} &
\colhead{Ephemeris F} &
\colhead{Ephemeris G} &
\colhead{Ephemeris H} &
\colhead{Ephemeris I}\\
\colhead{Parameter} &
\colhead{Spanning MJD} &
\colhead{Spanning MJD} &
\colhead{Spanning MJD} &
\colhead{Spanning MJD} &
\colhead{Spanning MJD} &
\colhead{Spanning MJD} &
\colhead{Spanning MJD} &
\colhead{Spanning MJD} &
\colhead{Spanning MJD}\\
&
\colhead{50826$-$51418} &
\colhead{51447$-$51996} &
\colhead{52036$-$52960} &
\colhead{53010$-$53325} &
\colhead{53377$-$53548} &
\colhead{53556$-$54541} &
\colhead{54548$-$54786} &
\colhead{54837$-$55517} &
\colhead{55568$-$55882}
}
\startdata
%%                                           & A              & B              & C                       & D               & E             & F               & G               & HFIX             & I\\
MJD start                                    & 50825.792      & 51446.610      & 52035.655               & 53010.094       & 53377.133     & 53555.734       & 54547.602       & 54836.798        & 55567.971\\
MJD end                                      & 51418.374      & 51995.679      & 52960.186               & 53325.061       & 53547.811     & 54540.671       & 54785.834       & 55516.933        & 55882.248\\
TOAs                                         & 39             & 20             & 74                      & 69              & 29            & 124             & 34              & 84               & 45\\
$\nu$ (Hz)                                   & 0.090913817(3) & 0.090906070(2) & 0.090906067(11)         & 0.090892729(14) & 0.09088760(2) & 0.090885255(16) & 0.090871541(12) & 0.0908675842(14) & 0.090857365(5)\\
$\dot{\nu}$ (10$^{-13}$ Hz s$^{-1}$)         & $-$1.582(5)    & $-$1.575(2)    & $-$1.556(6)             & $-$1.39(5)      & $-$1.17(9)    & $-$2.38(11)     & $-$1.50(5)      & $-$1.6064(11)    & $-$1.618(8)\\
$\ddot{\nu}$ (10$^{-22}$ Hz s$^{-2}$)        & $-$1.3(4)      & 3.2(8)         & $-$8.1(1.8)             & $-$45(11)       & $-$138(25)    & 400(50)         & $-$30(11)       & $-$0.88(4)       & $-$1.0(6)\\
$d^{3}\nu/dt^{3}$ (10$^{-28}$ Hz s$^{-3}$)   & $-$0.050(12)   & $-$            & 1.4(3)                  & 5.5(1.6)        & 16(2)         & $-$136(17)      & 3.0(1.0)        & $-$              & $-$\\
$d^{4}\nu/dt^{4}$ (10$^{-35}$ Hz s$^{-4}$)   & $-$            & $-$            & $-$1.3(3)               & $-$3.1(1.0)     & $-$           & 343(44)         & $-$             & $-$              & $-$\\
$d^{5}\nu/dt^{5}$ (10$^{-43}$ Hz s$^{-5}$)   & $-$            & $-$            & 6.9(1.5)                & $-$             & $-$           & $-$6153(864)    & $-$             & $-$              & $-$\\
$d^{6}\nu/dt^{6}$ (10$^{-50}$ Hz s$^{-6}$)   & $-$            & $-$            & $-$1.7(4)               & $-$             & $-$           & 8494(1300)      & $-$             & $-$              & $-$\\
$d^{7}\nu/dt^{7}$ (10$^{-54}$ Hz s$^{-7}$)   & $-$            & $-$            & $-$                     & $-$             & $-$           & $-$8.6(1.5)     & $-$             & $-$              & $-$\\
$d^{8}\nu/dt^{8}$ (10$^{-61}$ Hz s$^{-8}$)   & $-$            & $-$            & $-$                     & $-$             & $-$           & 6.1(1.1)        & $-$             & $-$              & $-$\\
$d^{9}\nu/dt^{9}$ (10$^{-68}$ Hz s$^{-9}$)   & $-$            & $-$            & $-$                     & $-$             & $-$           & $-$2.7(6)       & $-$             & $-$              & $-$\\
$d^{10}\nu/dt^{10}$ (10$^{-76}$ Hz s$^{-10}$) & $-$           & $-$            & $-$                     & $-$             & $-$           & 5.8(1.3)        & $-$             & $-$              & $-$\\
$\Delta \nu _{d}$\tablenotemark{b} (Hz)      & $-$            & $-$            & 36(3)$\times$10$^{-08}$ & $-$             & $-$           & $-$             & $-$             & $-$              & $-$ \\
$t_d$\tablenotemark{b} (days)                & $-$            & $-$            & 43(2)                   & $-$             & $-$           & $-$             & $-$             & $-$              & $-$ \\
Epoch (MJD)                                  & 51445.3846     & 52016.48413    & 52016.48413             & 52989.8475      & 53366.3150    & 53555.0000      & 54547.0000      & 54836.0000       & 55567.0000\\
RMS residual (phase)                         & 0.011          & 0.012          & 0.015                   & 0.013           & 0.016         & 0.019           & 0.022           & 0.022?          & 0.026\\
\enddata
\tablenotetext{a} {Numbers in parentheses are TEMPO-reported 1$\sigma$
uncertainties. }
\tablenotetext{b}{Parameters held fixed at values determined from the glitch fit
shown in Table~\ref{TBIG}.}
\end{deluxetable}

% >>>>>>>>>>>>>>>>>>>>>>>>>>>>>>>>>>>>>>>>>>>>>>> \ref{table-2259}

\clearpage
\begin{deluxetable}{lccccc}
\tabletypesize{\scriptsize}
%% \tabletypesize{\tiny}
\tablewidth{480.0pt}
%% \rotate
\tablecaption
{
Long-Term Spin Parameters for 1E~2259+586\tablenotemark{a}
\label{table-2259}
}
\tablehead
{
&
\colhead{Ephemeris A} &
\colhead{Ephemeris B1\tablenotemark{b}} &
\colhead{Ephemeris B2\tablenotemark{b}} &
\colhead{Ephemeris C} &
\colhead{Ephemeris D}\\
\colhead{Parameter} &
\colhead{Spanning MJD} &
\colhead{Spanning MJD} &
\colhead{Spanning MJD} &
\colhead{Spanning MJD} &
\colhead{Spanning MJD}\\
&
\colhead{50356$-$52398} &
\colhead{52445$-$52462} &
\colhead{52503$-$54181} &
\colhead{54194$-$54852} &
\colhead{54852$-$55924}
}
\startdata
%%                                        & A               & B1             & B2              & C                & D
MJD start                                 & 50355.891       & 52445.056      & 52503.050       & 54194.198        & 54852.303\\
MJD end                                   & 52398.421       & 52461.975      & 54180.566       & 54852.305        & 55924.287\\
TOAs                                      & 99              & 10             & 206             & 99               & 140\\
$\nu$ (Hz)                                & 0.1432887919(2) & 0.143287611(6) & 0.1432874928(2) & 0.14328612668(3) & 0.14328554678(3)\\
$\dot{\nu}$ (10$^{-14}$ Hz s$^{-1}$)      & $-$1.0201(17)   & $-$4.8(7)      & $-$0.9729(13)   & $-$0.99307(13)   & $-$0.96748(7)\\
$\ddot{\nu}$ (10$^{-24}$ Hz s$^{-2}$)     & 5.4(5)          & $-$            & $-$6.5(4)       & $-$              & $-$\\
$d^{3}\nu/dt^{3}$ (10$^{-31}$ Hz s$^{-3}$) & $-$0.43(7)     & $-$            & 1.05(5)         & $-$              & $-$\\
Epoch (MJD)                               & 50355.0000      & 52445.0000     & 52445.0000      & 54187.0000       & 54852.0000\\
RMS residual (phase)                      & 0.016           & 0.0077         & 0.0087          & 0.012            & 0.010\\
\enddata\
\tablenotetext{a} {Numbers in parentheses are TEMPO-reported 1$\sigma$
uncertainties. }
\tablenotetext{b} {The B ephemeris is split into two segments.
The first segment excludes the observation containing bursts, and
stops at the end of the recovery of the pulsar from a large glitch.}
\end{deluxetable}

% >>>>>>>>>>>>>>>>>>>>>>>>>>>>>>>>>>>>>>>>>>>>>>> \ref{table-0142}

\clearpage
\begin{deluxetable}{lccccc}
\tabletypesize{\scriptsize}
%% \tabletypesize{\tiny}
\tablewidth{470.0pt}
%% \rotate
\tablecaption
{
Long-Term Spin Parameters for 4U~0142+61\tablenotemark{a}
\label{table-0142}
}
\tablehead
{
&
\colhead{Ephemeris A} &
\colhead{Ephemeris B} &
\colhead{Ephemeris C} &
\colhead{Ephemeris D} &
\colhead{Ephemeris E}\\
\colhead{Parameter} &
\colhead{Spanning MJD} &
\colhead{Spanning MJD} &
\colhead{Spanning MJD} &
\colhead{Spanning MJD} &
\colhead{Spanning MJD}\\
&
\colhead{50171$-$50893} &
\colhead{51611$-$53800} &
\colhead{53831$-$54867} &
\colhead{54881$-$55763} &
\colhead{55778$-$55917}
}
\startdata
%%                         
%%                                        & A               & B               & C               & D                & E                 
MJD start                                 & 50170.693       & 51610.636       & 53831.335       & 54881.291        & 55777.699\\
MJD end                                   & 50893.288       & 53800.134       & 54867.298       & 55762.825        & 55917.361\\
TOAs                                      & 19              & 118             & 116             & 70               & 11\\
$\nu$ (Hz)                                & 0.115096868(3)  & 0.1150921156(7) & 0.1150921068(7) & 0.1150920514(11) & 0.115088121(5)\\
$\dot{\nu}$ (10$^{-14}$ Hz s$^{-1}$)      & $-$2.659(3)     & $-$2.679(5)     & $-$2.714(4)     & $-$2.621(8)      & $-$2.64(6)\\
$\ddot{\nu}$ (10$^{-23}$ Hz s$^{-2}$)     & $-$             & $-$2.0(2)       & 1.14(8)         & $-$              & $-$\\
$d^{3}\nu/dt^{3}$ (10$^{-31}$ Hz s$^{-3}$) & $-$            & $-$5.5(5)       & $-$             & $-$              & $-$\\
$d^{4}\nu/dt^{4}$ (10$^{-39}$ Hz s$^{-4}$) & $-$            & $-$5.5(5)       & $-$             & $-$              & $-$\\
Epoch (MJD)                               & 51704.000025    & 53800.0000      & 53800.0000      & 53800.0000       & 55762.0000\\
RMS residual (phase)                      & 0.019           & 0.015           & 0.021           & 0.016            & 0.012\\
\enddata
\tablenotetext{a} {Numbers in parentheses are TEMPO-reported 1$\sigma$
uncertainties. }
\end{deluxetable}

% >>>>>>>>>>>>>>>>>>>>>>>>>>>>>>>>>>>>>>>>>>>>>>> \ref{TBIG} GIANT TABLE

\clearpage

\pagestyle{empty}

%_____________________________________________________________________________
\begin{deluxetable}{lccccccccc}
\tablecolumns{10}
\tabletypesize{\tiny}
\tablewidth{700.0pt}
\rotate
\tablecaption{Glitches Observed in the Monitored Anomalous X-ray
Pulsars\tablenotemark{*,\dag} 
\label{TBIG}
}
%_____________________________________________________________________________
\tablehead
{
\colhead{Glitch} &
\colhead{MJD Range\tablenotemark{b}} &
\colhead{Glitch} &
\colhead{$\Delta\nu$\tablenotemark{c}} &
\colhead{$\Delta\nu/\nu$\tablenotemark{d}} &
\colhead{$\Delta\dot{\nu}$\tablenotemark{e}} &
\colhead{$Q$\tablenotemark{f} ~of } &
\colhead{$\tau$\tablenotemark{f} ~of } &
\colhead{Associated} &
\colhead{References} \\
\colhead{Number\tablenotemark{a}} &
&
\colhead{Epoch} &
&
&
&
\colhead{Recovery} &
\colhead{Recovery} &
\colhead{Radiative} &
\\
&
&
\colhead{(MJD)} &
\colhead{(Hz)} &
&
\colhead{(Hz/s)} &
&
\colhead{(days)} &
\colhead{Events\tablenotemark{g}} &
\\
}
\startdata
%_____________________________________________________________________________
%% \cutinhead{AXP 1E~1841$-$045, $P$~$\sim$~11.78~s, $\nu$~$\sim$~0.084890~Hz, 
%% $\dot{\nu}$\tablenotemark{g}~$\sim$~$-$2.95$\times$10$^{-13}$~Hz/s,
%% B\tablenotemark{h}~$\sim$~7.0$\times$10$^{14}$~G.}

\cutinhead{{1E 1841$-$045}}

%% $$ & $$ & $$ & $$ & $$ & $$ & $$ & $$ & $$ \\

Glitch~1\tablenotemark{h} & 52300$-$52610 & 52453.132194 & 2.91(10)$\times$10$^{-7}$ & 
3.43(12)$\times$10$^{-6}$ & $-$1.28(12)$\times$10$^{-14}$ & --- & --- & --- &
\cite{dkg08} \\

$$ & $$ & $$ & $$ & $$ & $$ & $$ & $$ & $$ & \cite{RIMASPEN} \\
$$ & $$ & $$ & $$ & $$ & $$ & $$ & $$ & $$ & \cite{RIMTHESIS} \\

Glitch~2 & 52773$-$53244 & 52997.049200 & 2.08(4)$\times$10$^{-7}$ & 
2.45(4)$\times$10$^{-6}$ & +4(3)$\times$10$^{-16}$ & ---&  --- & --- & \cite{dkg08} \\

Glitch~3 & 53579$-$53971 & 53823.969400 & 1.17(7)$\times$10$^{-7}$ & 
1.38(9)$\times$10$^{-6}$ & +2(1)$\times$10$^{-15}$ & --- & --- & --- & \cite{dkg08} \\

Glitch~4 & 54209$-$54348 & 54304.150312 & 4.6(3)$\times$10$^{-7}$ & 
5.5(4)$\times$10$^{-6}$ & $-$2.1(1.4)$\times$10$^{-14}$ & --- & --- & --- &
\cite{RIM40YEARS} \\

$$ & $$ & $$ & $$ & $$ & $$ & $$ & $$ & $$ &  \cite{RIMASPEN} \\
$$ & $$ & $$ & $$ & $$ & $$ & $$ & $$ & $$ &  \cite{RIMTHESIS} \\

Notable & 55363$-$55728 & 55596.958500 & 8.2(7)$\times$10$^{-8}$ & 
8.2(7)$\times$10$^{-8}$ & +4.4(1.2)$\times$10$^{-15}$ & --- & --- &
{\emph{Swift, Fermi}} \& {\emph{RXTE}} detected & This paper \\

Timing & $$ & $$ & $$ & $$ & $$ & $$ & $$ & bursts within a few days & $$ \\
Discontinuity\tablenotemark{i} & $$ & $$ & $$ & $$ & $$ & $$ & $$ & $$ & $$ \\

%_____________________________________________________________________________
%% \cutinhead{AXP RXS~J170849.0$-$400910, $P$~$\sim$~11.00~s, $\nu$~$\sim$~0.090893~Hz,
%% $\dot{\nu}$~$\sim$~$-$1.60$\times$10$^{-13}$~Hz/s, B~$\sim$~4.7$\times$10$^{14}$~G.}

\cutinhead{{RXS J170849.0$-$400910}}

Glitch~1 & 51186$-$51614 & 51445.384600 & 5.1(3)$\times$10$^{-8}$ & 
5.6(3)$\times$10$^{-7}$ & $-$8(4)$\times$10$^{-16}$ & --- &  --- & --- & \cite{klc00} \\

$$ & $$ & $$ & $$ & $$ & $$ & $$ & $$ & $$ & \cite{dkg08} \\

Glitch~2 & 51614$-$52366 & 52016.484130 & 2.2(4)$\times$10$^{-8}$ (long-term) & 
4.2(2)$\times$10$^{-6}$ & $-$1.1(2)$\times$10$^{-15}$ & 0.94(11) & 43(2) 
& Possible profile changes & \cite{kg03} \\

$$ & $$ & $$ & +3.6(3)$\times$10$^{-7}$ (recovered) & $$ & $$ & $$ & $$ & $$ & \cite{dis+03} \\
$$ & $$ & $$ & $$ & $$ & $$ & $$ & $$ & $$ & \cite{dkg08} \\

%% %Candidate- & 52745$-$53140 & 52989.847500 & 2.8(4)$\times$10$^{-8}$ & 
%% %3.1(5)$\times$10$^{-7}$ & 0\tablenotemark{j} & --- & --- & \cite{dkg08} \\

%% %glitch~1\tablenotemark{k,o} & $$ & $$ & $$ & $$ & $$ & $$ & $$ & $$ \\

%% %Discontinuity & 52745$-$53140 & 52989.847500 & 2.8(4)$\times$10$^{-8}$ &
%% %3.1(5)$\times$10$^{-7}$ & 0\tablenotemark{XX} & --- & --- & \cite{dkg08} \\

%% %$$ & $$ & $$ & $$ & $$ & $$ & $$ & $$ & $$ \\

Candidate- & 53229$-$53456 & 53366.315000 & 5.5(8)$\times$10$^{-8}$ & 
6.0(8)$\times$10$^{-7}$ & $-$1.9(1.3)$\times$10$^{-15}$\tablenotemark{l}
& --- & --- & Possible profile changes & \cite{dkg08} \\

glitch~1\tablenotemark{j,k} & $$ & $$ & $$ & $$ & $$ & $$ & $$ & $$ & \cite{igz+07} \\

Glitch~3 & 53465$-$53631 & 53549.150950 & 2.46(9)$\times$10$^{-7}$ & 
2.71(10)$\times$10$^{-6}$ & $-$2(2)$\times$10$^{-15}$\tablenotemark{m} & --- & --- & --- & \cite{dkg08} \\

$$ & $$ & $$ & $$ & $$ & $$ & $$ & $$ & $$ & \cite{igz+07} \\

%% %CHECKDiscontinuity & MJD-MJD & EPOCH & JUMP &
%% %JUMP & JUMP & --- & --- & This paper \\

%% %CHECKDiscontinuity & MJD-MJD & EPOCH & JUMP &
%% %JUMP & JUMP & --- & --- & This paper \\

Candidate- & 55315$-$55755 & 55517.1247 & 9.4(5)$\times$10$^{-8}$ & 
1.03(5)$\times$10$^{-6}$ & +1.4(4)$\times$10$^{-15}$ & --- & --- & --- & This paper \\

glitch~2\tablenotemark{j} & $$ & $$ & $$ & $$ & $$ & $$ & $$ & $$ & $$ \\

%_____________________________________________________________________________
%% \cutinhead{AXP 1E~2259+586, $P$~$\sim$~6.98~s, $\nu$~$\sim$~0.143286~Hz, 
%% $\dot{\nu}$~$\sim$~$-$0.0992$\times$10$^{-13}$~Hz/s,
%% B~$\sim$~0.59$\times$10$^{14}$~G.}

\cutinhead{{1E 2259+586}}

Glitch~1\tablenotemark{n} & 51623$-$52900 & 52443.130000 &
$\sim$4.95$\times$10$^{-7}$ (long-term) &  4.24(11)$\times$10$^{-6}$ &
+2.18(25)$\times$10$^{-16}$ & 0.185(10) & 12$-$17
& Bursts, pulsed and total & \cite{wkt+04} \\

$$ & $$ & $$ & +~$\sim$1.12$\times$10$^{-7}$ (recovered) &
$$ & $$ & $$ & $$ & flux enhancements, & \cite{kgw+03} \\

$$ & $$ & $$ & $$ &
$$ & $$ & $$ & $$ & large profile changes & $$ \\

%% on next line, slightly exagerated error on delnu
Glitch~2 & 54085$-$54256 & 54184.903200 & 1.261(4)$\times$10$^{-7}$ & 
8.80(3)$\times$10$^{-7}$ & $-$6(2)$\times$10$^{-16}$ & --- & --- & --- & \cite{dkg08} \\

$$ & $$ & $$ & $$ & $$ & $$ & $$ & $$ & $$ & \cite{RIM40YEARS} \\
$$ & $$ & $$ & $$ & $$ & $$ & $$ & $$ & $$ & \cite{RIMASPEN} \\
$$ & $$ & $$ & $$ & $$ & $$ & $$ & $$ & $$ & \cite{RIMTHESIS} \\
$$ & $$ & $$ & $$ & $$ & $$ & $$ & $$ & $$ & \cite{ibi12} \\

Notable & 54571$-$55112 & 54832$-$54880\tablenotemark{o} & $-$1.2(3)$\times$10$^{-8}$\tablenotemark{o} &
$-$8.2(2.1)$\times$10$^{-8}$ & +2.3(1.6)$\times$10$^{-16}$ & --- & --- &
Profile change in 1 obs. & \cite{ibi12} \\

Timing & $$ & $$ & $$ & $$ & $$ & $$ & $$ & Pulsed flux and count & This paper \\
Discontinuity\tablenotemark{o} & $$ & $$ & $$ & $$ & $$ & $$ & $$ & rate change in 2 obs. & $$ \\

%_____________________________________________________________________________
%% \cutinhead{AXP 4U~0142+61, $P$~$\sim$~8.69~s, $\nu$~$\sim$~0.115092~Hz,
%% $ \dot{\nu}$~$\sim$~$-$0.268$\times$10$^{-13}$~Hz/s,
%% B~$\sim$~1.3$\times$10$^{14}$~G.}

\cutinhead{{4U 0142+61}}

Candidate- & 50170$-$52340 & 50893$-$51610\tablenotemark{q} &
(6.7(3)$-$8.1(3))$\times$10$^{-8}$ & 
(5.8(2)-7.0(2))$\times$10$^{-7}$ & $-$2.4(3)$\times$10$^{-16}$ & --- & --- &
Possible profile changes & \cite{mks05} \\

glitch~1\tablenotemark{p} & $$ & $$ & $$ & $$ & $$ & $$ & $$ & $$ & \cite{dkg07} \\

Candidate- & 53481$-$54235 & 53809.185840 & $-$1.27(17)$\times$10$^{-8}$ (long-term) &
1.6(4)$\times$10$^{-6}$ & $-$3.1(1.2)$\times$10$^{-16}$ & 1.0(3) & 17.0(1.7) & Bursts, short-term 
& \cite{gdk11} \\

glitch~2\tablenotemark{r} & $$ & $$ & +2.0(4)$\times$10$^{-7}$ (recovered) &
$$ & $$ & $$ & $$ & pulsed flux increase\tablenotemark{s}, & $$ \\

$$ & $$ & $$ & $$ & $$ & $$ & $$ & $$ & pulse profile changes & $$ \\

Glitch~1\tablenotemark{t} $\;\;\;\;\;\;\;\;$ & 55329$-$55917 & 55771.190600 &
5.11(4)$\times$10$^{-7}$ & 4.44(4)$\times$10$^{-6}$ &
0\tablenotemark{t} & --- $\;\;\;\;$ & --- $\;\;\;\;$ & Hint of a pulsed flux & This paper \\

$$ & $$ & $$ & $$ & $$ & $$ & $$ & $$ & increase (1$\sigma$ level) & $$ \\

%_____________________________________________________________________________
%% \cutinhead{AXP 1E~1048.1$-$5937, $P$~$\sim$~6.46~s, $\nu$~$\sim$~0.1549~Hz,
%% $\dot{\nu}$~$\sim$~$-$5.33$\times$10$^{-13}$~Hz/s,
%% B~$\sim$~3.8$\times$10$^{14}$~G.}

\cutinhead{{1E 1048.1$-$5937}}

%% Candidate- & 52281$-$52485 & 52386.0(1.5) & 4.51(14)$\times$10$^{-7}$ &
%% 2.91(9)$\times$10$^{-6}$ &  $-$4.10(15)$\times$10$^{-14}$ & --- & Onset of second pulsed & \cite{dkg09} \\

%% glitch~1 & $$ & $$ & $$ & $$ & $$ & $$ & (and total) flux flare, & $$ \\

%% $$ & $$ & $$ & $$ & $$ & $$ & $$ & large profile changes & $$ \\

Glitch~1\tablenotemark{u,v} & 54164$-$54202 & 54185.912956 & 2.52(3)$\times$10$^{-6}$ & 
1.63(2)$\times$10$^{-5}$ & $-$6(4)$\times$10$^{-14}$ & --- & --- & Onset of third pulsed & \cite{dkg09} \\

$$ & $$ & $$ & $$ & $$ & $$ & $$ & $$ & (and total) flux flare, & \cite{RIM40YEARS} \\

$$ & $$ & $$ & $$ & $$ & $$ & $$ & $$ & some profile changes & \cite{tgd+08} \\

%_____________________________________________________________________________
\enddata
\tablenotetext{*}{Numbers in parentheses are {\texttt{TEMPO}}-reported 1$\sigma$
uncertainties.}%%
\tablenotetext{\dag}{See \S\ref{sec:tablenotes} for the text of the
tablenotes ``a'' to ``v''.}%%
%% %% \tablenotetext{\dagger}{Chapter~2: \cite{dkg07UPDATED}, Chapter~3: \cite{gdk09submitted},
%% %% Chapter~4: \cite{dkg08}, Chapter~5: \cite{dkg09submitted}.}%
%% %%
\end{deluxetable}

% >>>>>>>>>>>>>>>>>>>>>>>>>>>>>>>>>>>>>>>>>>>>>>> \ref{table-bursts} 

\clearpage
\begin{deluxetable}{lccccc}
\tabletypesize{\scriptsize}
\tablewidth{520.0pt}
%% \rotate
\tablecaption
{
{\emph{RXTE}}-Detected Bursts
\label{table-bursts}
}
\tablehead
{
Source &
Date &
Observation &
Number of &
Reference &
Associated \\
$$ &
$$ &
Containing &
Bursts &
$$ &
With \\
$$ &
$$ &
Bursts &
Detected &
$$ &
$$
}
\startdata
1E~1841$-$045 & 07/05/2010 & 95017-03-01-00 & 1 & This paper & The 2010 May period of\\
$$ & $$ & $$ & $$ & $$ & bursting activity, as observed\\
$$ & $$ & $$ & $$ & $$ & by {\emph{Swift}} and {\emph{Fermi}}, \citep{lkg+11}.\\
RXS~J170849.0$-$400910 & $-$ & $-$ & $-$ & $-$ & $-$ \\
1E~2259+586 & 18/06/2002 & 70094-01-03-00 & $>$80 & \cite{gkw04} & The 2002 July major outburst.\\
$$ & $$ & $$ & $$ & \cite{kgw+03} & $$ \\
$$ & $$ & $$ & $$ & \cite{wkt+04} & $$ \\
4U~0142+61 & 06/04/2006 & 92006-05-03-00 & 1 & \cite{gdk11} & The 2006$-$2007 active phase. \\ 
$$         & 25/06/2006 & 92006-05-09-01 & 4 & \cite{gdk11} & The 2006$-$2007 active phase. \\
$$         & 07/02/2007 & 92006-05-25-00 & 1 & \cite{gdk11} & The 2006$-$2007 active phase. \\
1E~1048.1$-$5937 & 29/10/2001 & 60069-03-35-00 & 1 & \cite{gkw02} & The first of two slow-rise flares.\\
$$ & $$ & $$ & $$ & \cite{gk04} & $$ \\
$$               & 14/11/2001 & 60069-03-37-00 & 1 & \cite{gkw02} & The first of two slow-rise flares.\\
$$ & $$ & $$ & $$ & \cite{gk04} & $$ \\
$$               & 29/06/2004 & 90076-02-09-02 & 1 & \cite{gkw06} & $-$\\
$$               & 04/04/2007 & 92005-02-01-00 & Possibly~1 & This paper & The 2007 outburst. \\
$$               & 28/04/2007 & 92005-02-08-01 & 1 & \cite{dkg09} & The 2007 outburst. \\
\enddata
%% \tablenotetext{a} {??}
\end{deluxetable}

% >>>>>>>>>>>>>>>>>>>>>>>>>>>>>>>>>>>>>>>>>>>>>>> \ref{table-summary} 

\clearpage
\begin{deluxetable}{llccc}
\tabletypesize{\footnotesize}
\tablewidth{400.0pt}
%% \rotate
\tablecaption
{
Summary of {\it RXTE}-observed AXP Notable Event Phenomenology
\label{table-summary}
}
\tablehead
{$$ &
Accompanying &
Accompanying &
Accompanying &
Accompanying \\
Phenomenon &
Glitch/Anomaly\tablenotemark{a} &
Flux Increase\tablenotemark{b} &
Profile Change\tablenotemark{c} &
X-ray Bursts\tablenotemark{d} }
\startdata
Glitch/Anomaly\tablenotemark{a} & {\ldots}  & 5/22 & 6/22 & 5/22 \\
Flux Increase\tablenotemark{b} &  5/5 & {\ldots}  & 5/5 & 4/5 \\
Profile Change\tablenotemark{c} & 6/6 & 5/6 & {\ldots}  & 5/6 \\
X-ray Bursts\tablenotemark{d} & 5/6 & 4/6 & 5/6 & {\ldots}  \\
\enddata
\tablenotetext{a} {The number of any form of {\emph{RXTE}}-observed
timing anomaly including glitches, and excluding the early 2012 event
for AXP~1E~1048.1$-$5937, which occured within days of our last
observation.}
\tablenotetext{b} {The number of {\emph{RXTE}}-observed long-lived
episodes of pulsed flux variations,
excluding the small hint of pulsed flux increase in
AXP~4U~0142+61 following its 2011 glitch.}
\tablenotetext{c} {The number of {\emph{RXTE}}-observed pulse profile
changes, excluding the two possible pulse profile changes
following timing anomalies in AXP~RXS~J170849.0$-$400910.}
\tablenotetext{d} {The number of {\emph{RXTE}}-observed X-ray burst
events. When a timing anomaly or a slow radiative event were
followed by several bursts, the collection of bursts counts as one
burst event.}
\end{deluxetable}
\clearpage

%% --------------------------------------------------------
%% --------------------------------------------------------
%% --------------------------------------------------------
%% --------------------------------------------------------

\bibliographystyle{apj}

%% --------------------------------------------------------
\end{document}